\documentclass[10pt,twocolumn,twoside]{IEEEtran}
\usepackage{amssymb,amsmath,epsfig,cite,comment,hyperref,verbatim,graphicx,multirow,bm}

\DeclareMathOperator*{\argmin}{arg\,min}
\usepackage{setspace,url}
\usepackage{xcolor}
\usepackage{soul}

\title{Time-varying quasi-closed-phase analysis for accurate formant tracking in speech signals}
\author{Dhananjaya Gowda~(\IEEEmembership{Member,~IEEE}), Sudarsana Reddy Kadiri$^*$~(\IEEEmembership{Member,~IEEE}), Brad Story, and \\ Paavo Alku~(\IEEEmembership{Fellow,~IEEE})
\thanks{$^*$Corresponding author. \newline D. Gowda was with Aalto University at the time this work was carried out, and he is now with Samsung Research Korea, E-mail: njaygowda@gmail.com. S.R. Kadiri and P. Alku are with the Dept. of Signal Processing and Acoustics, Aalto University, Finland, E-mail: \{sudarsana.kadiri, paavo.alku\}@aalto.fi. B. Story is with the University of Arizona, USA. This study was funded by the Academy of Finland (project no. 284671, 312490).}
}

\begin{document}
\maketitle
\begin{abstract}
In this paper, we propose a new method for the accurate estimation and tracking of formants in speech signals using time-varying quasi-closed-phase (TVQCP) analysis.
Conventional formant tracking methods typically adopt a two-stage estimate-and-track strategy wherein an initial set of formant candidates are estimated using short-time analysis (e.g., 10--50 ms), followed by a tracking stage based on dynamic programming or a linear state-space model.
One of the main disadvantages of these approaches is that the tracking stage, however good it may be, cannot improve upon the formant estimation accuracy of the first stage.
The proposed TVQCP method provides a single-stage formant tracking that combines the estimation and tracking stages into one. TVQCP analysis combines three approaches to improve formant estimation and tracking: (1) it uses temporally weighted quasi-closed-phase analysis to derive closed-phase estimates of the vocal tract with reduced interference from the excitation source, (2) it increases the residual sparsity by using the $L_1$ optimization and (3) it uses time-varying linear prediction analysis over long time windows (e.g., 100--200 ms) to impose a continuity constraint on the vocal tract model and hence on the formant trajectories.
Formant tracking experiments with a wide variety of synthetic and natural speech signals show that the proposed TVQCP method performs better than conventional and popular formant tracking tools, such as Wavesurfer and Praat (based on dynamic programming), the KARMA algorithm (based on Kalman filtering), and DeepFormants (based on deep neural networks trained in a supervised manner). Matlab scripts for the proposed method can be found at: \url{https://github.com/njaygowda/ftrack}
\end{abstract}


\section{Introduction}
\label{sec:intro}
Vocal tract resonances (VTRs), commonly referred to as {\em formant frequencies}, are speech parameters that are of fundamental importance in all areas of speech science and technology. The estimation and tracking of VTRs from speech signals is a challenging problem that has many applications in various areas: in acoustic and phonetic analysis \cite{fant1960, Assmann1995}, in voice morphing \cite{Rita2016}, in speech recognition \cite{Welling1998,Smit2012}, in speech and singing voice synthesis \cite{Pinto1989,Chan2015}, in voice activity detection \cite{Yoo2015}, and in designing hearing aids \cite{Schilling1998,Bruce2004}.
Many algorithms of varying complexity have been proposed in the literature for tracking formants in speech signals~\cite{wavesurfer2000, praat2001, Deng2007, Mehta2012, Durrieu2013}.
A dynamic programming (DP)--based tracking algorithm with a heuristic cost function on the initial formant candidates estimated using conventional linear prediction (LP) analysis was used in ~\cite{wavesurfer2000, praat2001}.
This two-stage approach has a detection stage, where an initial estimate of the VTRs is obtained, followed by a tracking stage.
An integrated approach towards tracking was adopted in ~\cite{Deng2007, Mehta2012, Durrieu2013} using state-space methods such as Kalman filtering (KF) and the factorial hidden Markov model (FHMM).
In both approaches, analysis of the signal for the accurate estimation (or modeling) of the vocal tract system is an important and necessary computational block.
However, it should be mentioned here that there are a few exceptions, such as \cite{Durrieu2013}, which uses a non-negative matrix factorization (NMF)--based source-filter modeling of speech signals. Recently, deep learning--based techniques \cite{Dissen2016,schiel2018,Dissen2019} have also been studied as alternatives to conventional statistical signal processing--based formant estimation and tracking methods. These methods, however, are based on supervised machine learning, which calls for having annotated speech corpora with which to obtain the ground truth formant frequencies for system training.

LP analysis is one of the most widely used methods for estimating VTRs from speech signals \cite{Atal1967,Itakura1968,makhoul1975}.
To improve the accuracy of LP, several variants of this all-pole spectral modeling method have been proposed \cite{Kay1988}.
Among the different modifications, autocorrelation and covariance analyses are the most popular LP methods in formant estimation and tracking \cite{praat2001,wavesurfer2000}.
Covariance analysis is known to give more accurate formant estimates than autocorrelation analysis, but the stability of the resulting all-pole filter is not guaranteed in covariance analysis  \cite{makhoul1975,Wong1979}.
Even though the filter instability must be avoided in applications where the signal needs to be reconstructed (such as speech synthesis and coding), the instability in itself is not a serious problem in formant tracking.
Compared to covariance analysis, closed-phase analysis is known to provide even more accurate VTR estimates by avoiding the open-phase regions of the glottal cycle, which are influenced by the coupling of the vocal tract with the trachea \cite{Steiglitz1977,Yegna1998}.
Closed-phase analysis, however, works better for utterances such as those of low-pitched male voices, which have more samples in the closed phase of the glottal cycle compared to high-pitched female and child voices that might have just a few closed-phase samples per glottal cycle.

As a remedy for the lack of data samples in formant estimation, a selective prediction of speech samples can be conducted in spectral modeling. A sample-selective prediction is used in weighted linear prediction (WLP) methods by giving different temporal weighting to the prediction error at each discrete time instant \cite{Mizoguchi1982,Yanagida1985,ChinHui1988,ma1993,magi2009,JouniIS2010,PaavoJASA2013,Manu2014}.
One such method, called sample selective linear prediction (SSLP) analysis, was proposed in \cite{Mizoguchi1982} for better modeling of the vocal tract area function. In SSLP, a hard rejecting weighting function is used to eliminate outlier samples in sample selection.
A more generalized WLP algorithm was developed in \cite{Yanagida1985} with a continuous weighting function for the prediction residual.
In \cite{ChinHui1988}, an iterative LP algorithm, robust linear prediction was proposed by utilizing the non-Gaussian nature of the excitation signal to derive a temporal weighting function based on the magnitude of the residual samples.

To improve the robustness of linear predictive spectrum estimation, a simple non-iterative WLP method was studied in  \cite{ma1993} based on the short-time energy (STE) weighting function. The STE weighting function is a temporal energy function that is computed, for example, in 1--2 ms frames of the speech signal waveform. The STE weighting function emphasizes the importance of the high-energy regions within a glottal cycle in computing the autocorrelation (or covariance) matrix. Therefore, this WLP method is similar to closed-phase LP analysis because the high-energy sections of voiced speech emphasized by the STE weighting correspond roughly to glottal closed-phase regions.
Since the publication of WLP in \cite{ma1993}, several variants of this all-pole modeling method have been developed and used, for example, in the robust feature extraction of speech \cite{ma1993,JouniIS2010} and in glottal inverse filtering (GIF) \cite{Manu2014,Achuth2019}. Some of these more recent WLP algorithms have also addressed the stability of the all-pole filter \cite{magi2009,PaavoJASA2013}. 
In \cite{PaavoJASA2013}, a new weighting function, called the attenuated main excitation (AME) window, was studied to improve the accuracy of formant estimation, especially for high-pitched voices. The AME function is designed to attenuate the effect of prominent speech samples in the vicinity of the glottal closure instants (GCIs) on the autocorrelation function. This is justified because these high-energy speech samples are greatly contributed to by the glottal source, which results in distortion of the formant estimates by the biasing effect of the glottal source. As a sequel to using AME as a temporal weighting function in WLP, the quasi-closed-phase (QCP) analysis of speech signals  was proposed in \cite{Manu2014} for the estimation of glottal flow with GIF.
QCP analysis uses a more generalized version of the AME weighting function, for example, with slanted edges instead of vertical ones. In addition, the weighting function of QCP analysis puts more emphasis on the closed-phase regions compared to the open-phase regions that are prone to subglottal coupling.
However, the previous experiments with QCP analysis in \cite{Manu2014} focused solely on GIF analysis of the voice source, without any evaluation of the QCP algorithm's performance in formant detection and estimation.

The spectral modeling of speech is conducted using conventional LP in short-time segments (5--50 ms) by assuming speech to be a quasi-stationary process~\cite{makhoul1975}. This traditional short-time analysis models the real, continuously varying human vocal tract system in a piecewise manner. In addition, the conventional methods based on short-time LP analysis typically use a two-stage detect-and-track approach in tracking formants ~\cite{wavesurfer2000,praat2001}.
It should be noted that even those formant tracking methods that directly track formants from the cepstral coefficients use this piecewise approximation of the vocal tract system~\cite{Deng2007, Mehta2012}.
In order to take into account the inherent slowness of the real human vocal tract (i.e., the system being inertial), time-varying linear prediction (TVLP) provides an attractive method that models the vocal tract over longer time-spans by defining the model parameters as a function of time by using selected, low-order basis functions~\cite{HallSP1983,SchnellIcassp2008,ChetupalliTVSIcassp2014}.

The solution to conventional LP involves minimizing the $L_2$ norm of the prediction error signal, the residual, with an inherent assumption that the excitation source signal is a Gaussian process~\cite{WongASSP1979,Kay1988}.
Based on the theory of compressed sensing, sparsity constraints can be used to utilize the super Gaussian nature of the excitation signal~\cite{GiacobelloASLP2012,WipfIEEEJSTSP2010}.
This is achieved by approximating a non-convex $L_0$ norm optimization problem by using a more tractable convex $L_1$ norm optimization~\cite{GiacobelloASLP2012}.
In addition, it was shown in ~\cite{WipfIEEEJSTSP2010} that an iterative reweighted minimization of the norm can achieve increased sparsity of the error signal, which yields a solution closer to $L_0$ norm optimization.

In this article, we propose a new time-varying quasi-closed-phase (TVQCP) linear prediction analysis of speech for accurate modeling and tracking of VTRs. The proposed method aims to improve the estimation and tracking of formants by combining three different ideas: QCP analysis, increased sparsity of the error signal and time-varying filtering. To the best of our knowledge, this combination has not been studied before in formant estimation and tracking and is justified as follows. First, in order to reduce the effect of the glottal source in formant estimation, it is justified to take advantage of QCP analysis to temporally weight the prediction error, which has been shown to improve the estimation of the vocal tract in voice source analysis \cite{Manu2014,Achuth2019}. Second, filter optimization in previous QCP studies has been conducted using the $L_2$ norm which is known to result in less sparse residuals. Therefore, in order to further enhance the performance of temporal weighting, it is justified to increase the sparsity of the residual in QCP analysis by using the $L_1$ norm. Third, in order to take into account the fact that the natural human vocal tract is a slowly varying physiological system, we argue that formant tracking can be further improved by implementing the proposed $L_1$ norm -based QCP analysis using time-varying filtering. A preliminary investigation of TVQCP for formant tracking was published in a conference paper in ~\cite{Gowda2016}. In the current study, our preliminary experiments reported in ~\cite{Gowda2016} are expanded in many ways by, for example, including a larger number of evaluation datasets and a larger number of reference methods. In summary, the contributions of the current study are as follows.

\begin{itemize}
    \item Combining the ideas of QCP analysis, $L_1$ norm optimization and TVLP analysis to create a new formant estimation and tracking method, TVQCP. 
    \item Studying the advantages of sparsity by comparing the $L_1$ and $L_2$ norm optimization in TVQCP. 
    \item Analysing the effects of the different parameters in TVQCP.
    \item Studying the formant tracking performance of TVQCP using synthetic vowels of varying fundamental frequency values and phonation types, using high-pitched child speech simulated with a physical modeling approach, and using natural speech.
    \item Comparing TVQCP with popular formant tracking methods (Wavesurfer, Praat and KARMA) and with a recently proposed deep neural network -based method (DeepFormants) that is based on supervised learning. 
    \item Studying the noise robustness of TVQCP for different noise types and signal-to-noise ratio (SNR) scenarios. 
\end{itemize}

In the following two sections, the optimization of the TVQCP model is described by first presenting the time-invariant (i.e., stationary) QCP analysis in Section~\ref{sec:qcp} as background information. After this, the TVQCP (i.e., non-stationary QCP) analysis is presented in Section~\ref{sec:tvqcp}. Formant tracking experiments are reported in Section~\ref{sec:expt} and conclusions are drawn in Section~\ref{sec:summary}.

\section{Quasi-closed-phase analysis}
\label{sec:qcp}
QCP analysis belongs to the family of temporally weighted LP methods with a specially designed weighting function based on the knowledge of GCIs \cite{Manu2014}.
An overview of WLP and the design of the QCP weighting function is given in this section.

\subsection{Weighted linear prediction}
\label{sec:wlp}
In conventional LP, the current speech sample $x[n]$ is predicted based on the past $p$ speech samples as
\begin{equation}
\hat{x}[n]=-\sum_{k=1}^{p}{a_k x[n-k]} ,\label{eq:eq1}
\end{equation}
where $\{a_k\}_{k=0}^{p}$ with $a_0=1$ denote the prediction coefficients and $p$ is the prediction order.
Let us denote the estimated transfer function of the vocal tract system as $H(z)=1/A(z)$, where $A(z)$ is the $z-$transform of the prediction coefficients $\{a_k\}_{k=0}^{p}$.
The optimal prediction coefficients minimize the overall prediction error given by the cost function
\begin{gather}
E = \sum_{n}{e^2[n]},
\end{gather}
where $e[n]=x[n]-\hat{x}[n]$ is the sample-wise prediction error, the residual.
The optimal prediction coefficients are computed by minimizing the cost function ($\partial{E}/\partial{a_i}=0, \enskip 1\le i\le p$), which results in the following normal equations
\begin{gather}
\sum_{k=1}^{p}r_{i,k}a_k=-r_{i,0}, \quad 1\le i\le p ,\\
\text{where} \enskip r_{i,k}=\sum_{n}x[n-i]x[n-k].
\end{gather}

In the above formulation, it can be seen that the prediction error is minimized in the least-square sense by having equal temporal weighting for every sample.
However, in WLP, a different (positive) weighting value is imposed on each squared residual sample, resulting in the following WLP cost function
\begin{equation}
E_w = \sum_{n}{w[n] e^2[n]},
\end{equation}
where $w[n]$ denotes the weighting function on the sample-wise prediction error $e[n]$.
It should be noted that the weighting in WLP methods is on the error signal and should not be confused with the traditional short-time windowing (e.g., Hamming) of the speech signal that is used for reducing truncation effects in spectral analysis.
The prediction coefficients can be computed in a similar way to that of conventional LP by minimizing the cost function ($\partial{E_w}/\partial{a_i}=0, \enskip 1\le i\le p$) and solving the resulting normal equations
\begin{gather}
\sum_{k=1}^{p}b_{i,k}a_k=-b_{i,0}, \quad 1\le i\le p, \label{eq:eq6}\\
\text{where} \enskip b_{i,k}=\sum_{n}w[n]x[n-i]x[n-k].
\end{gather}

\subsection{The choice of weighting function}
\label{sec:wt_fn}
\begin{figure}[t]
\centering
\includegraphics[width=\columnwidth,trim=0.8cm 0.7cm 1cm 0.4cm,clip]{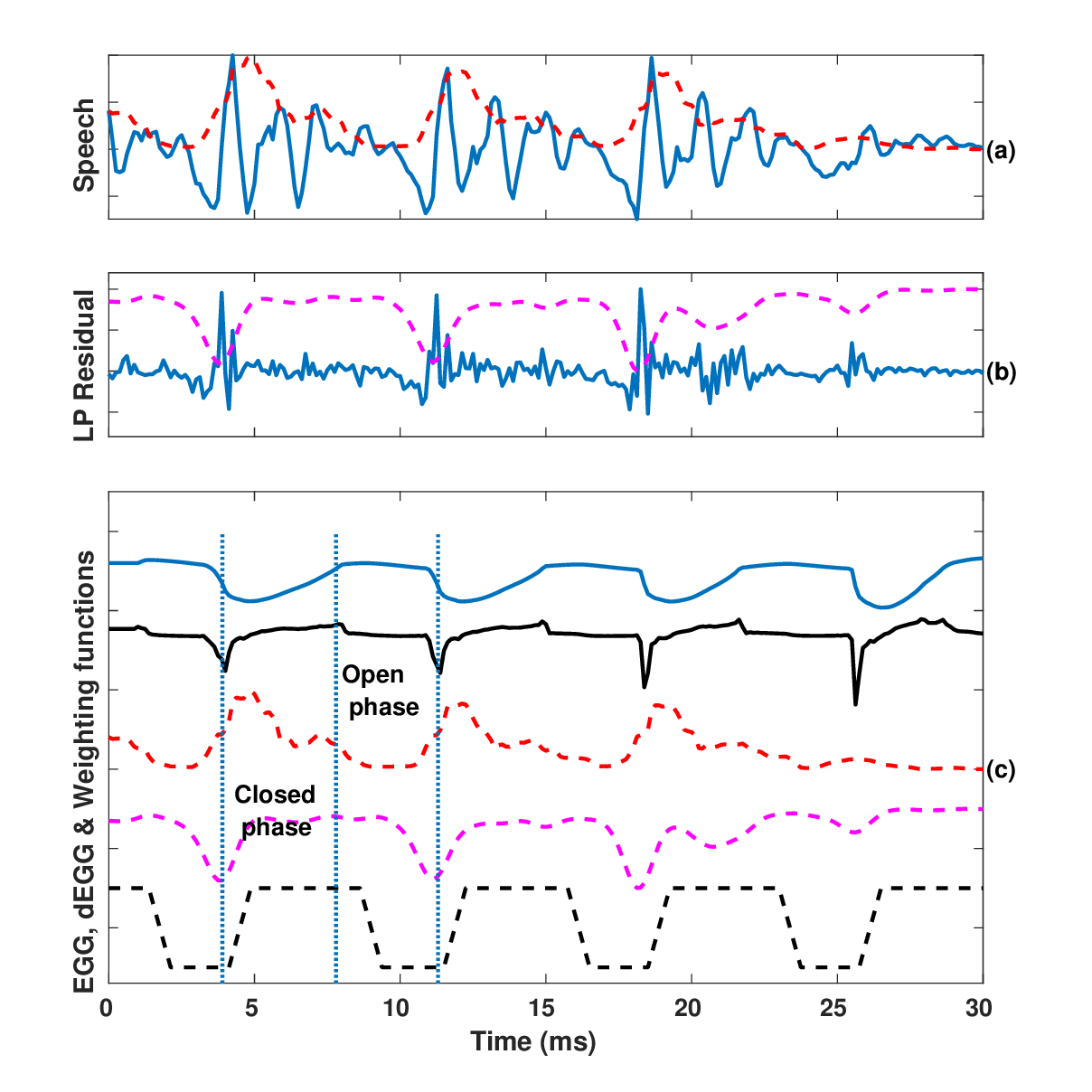}
\vspace*{-0.6cm}
\caption{\label{fig:ste} An illustration of different weighting functions for use in WLP: (a) the speech signal (the solid line) and the short-time energy (STE) weighting function (the dashed line); (b) the LP residual (the solid line) and the weighting function (the dashed line) derived from the residual; and (c) EGG (the solid blue line), dEGG (the solid black line), and three different weighting functions: speech signal--based STE weighting (the dashed red line), residual based weighting (the dashed pink line), and QCP weighting (the dashed black line).}
\end{figure}

As mentioned earlier in Section~\ref{sec:intro}, several weighting functions have been proposed for WLP.
STE is one of the popular weighting functions used in WLP, and it is demonstrated in Fig.~\ref{fig:ste}.
The figure shows an example of a vowel utterance, an electroglottography (EGG) signal, and the derivative of the EGG signal (dEGG), along with rough markings for the closed phases and open phases.
The STE weighting function is computed as
\begin{gather}
w[n]=\sum_{k=(D+1)}^{(D+M)}{x^2[n-k]},
\end{gather}
where the delay parameter $D$ controls the peak position (or emphasis) of the weighting function within the glottal cycle and
the length parameter $M$ controls the peak width, as well as the dynamic range and smoothness of the function. Typical values for these two parameters are $D=0$ and $M=12$, the latter corresponding to 1.5 ms at an 8 kHz sampling rate.
It can be seen that the STE function puts more weighting to the high-energy closed-phase regions of the glottal cycle.
However, Fig.~\ref{fig:ste} also demonstrates that the degree of suppression in both the glottal open phase and at the instant of the main excitation depends on the decay of the speech signal waveform within the glottal cycle. Therefore, the STE weighting function does not necessarily suppress these regions completely. The effect of this problem of the STE weighting function was studied in our previous study on formant estimation of high-pitched vowels \cite{PaavoJASA2013}. This previous study indicated that by changing the weighting function from STE to AME resulted in a clear improvement in formant estimation accuracy particularly for the first formant for which the average estimation accuracy improved by almost 10 percentage units.

A weighting function based on the residual signal energy can also be used. Fig.~\ref{fig:ste} shows a residual weighting function derived by inverting and normalizing (between 0 to 1) a zero-mean residual energy signal, computed similar to the STE function.
As can be seen from the figure, the residual weighting function may not suppress some weaker glottal excitations (at around 25 ms) as well as the stronger ones.
This effect can be more pronounced in the vowel beginning and ending frames with a highly transient signal energy.
Also, the residual weighting function may not effectively down-weight the contributions from the open-phase regions of the glottal cycle.
A QCP weighting function derived from knowledge of GCIs is also shown in Fig.~\ref{fig:ste}.
It can be seen that this weighting function emphasizes the closed-phase region of the glottal cycle, while at the same time the function de-emphasizes the region immediately after the main excitation as well as the open-phase region.

\subsection{Quasi-closed-phase weighting function}
\label{sec:qcpwt}
\begin{figure}[t]
\centering
\includegraphics[width=\columnwidth,trim=0.2cm 0.2cm 0 0,clip]{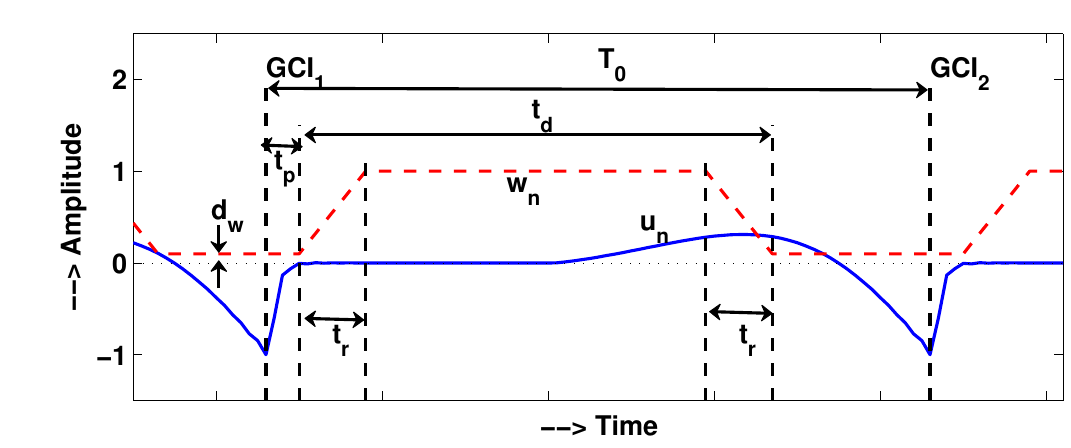}
\vspace*{-0.5cm}
\caption{\label{fig:amewin} The design of the quasi-closed-phase (QCP) weighting function $w_{n}$ (the dotted line), along with the LF glottal flow derivative signal $u_{n}$ (the solid line) for about one glottal cycle.}
\end{figure}

An example of the QCP weighting function $w_n$ is shown in Fig.~\ref{fig:amewin}, along with the Liljencrants-Fant (LF) glottal flow derivative waveform $u_n$ for about one glottal cycle. The QCP weighting function can be expressed with three parameters: the position quotient ($PQ=t_{p}/T_0$), the duration quotient ($DQ=t_{d}/T_0$), and ramp duration $t_{r}$, where $T_0$ is the time-length of the glottal cycle.
In order to avoid possible singularities in the weighted correlation matrix given in Eq. (\ref{eq:eq6}), a small positive value, $d_w=10^{-5}$, is used (instead of zero) as the minimum value in the weighting function.

The parameters of the QCP weighting function were optimized in \cite{Manu2014} using a set of about 65000 LF-excited synthetic vowels of different phonation types and different fundamental frequency values. Rather than aiming at a generic optimal weighting function, the optimization procedure adopted in \cite{Manu2014} was based on using a simple, pre-defined waveform depicted in Fig.~\ref{fig:amewin} whose parameters were optimized in a grid search. For more details about the optimization procedure, the reader is referred to Section IV.A in \cite{Manu2014}. The optimization procedure reported in \cite{Manu2014} gave both fixed QCP parameters and parameters where one of the values (DQ) was pitch-adaptive. In the current study, we used the pitch-adaptive QCP parameters of \cite{Manu2014} and the values of the two fixed parameters were as follows: $PQ$=0.05 and $t_{r}$=0.375 ms (which corresponds to $N_{ramp}$=3 samples using the notation of \cite{Manu2014}). $DQ$ was varied between 0.5 and 0.9 (as will be reported in Section~\ref{Robustness_to_GCI_DQ}) and was set to $DQ$=0.8.

Using the QCP function as a temporal weighting waveform in WLP provides two distinct advantages when compared to conventional LP (i.e., giving equal weighting to all squared residual samples) or conventional WLP (i.e., weighting is given using the STE function).
The first advantage is that the emphasis of the QCP weighting function is on the closed phase region, which provides more accurate modeling of the vocal tract by reducing the effect of coupling between subglottal and supraglottal cavities. The second is that the QCP weighting de-emphasizes the region immediately after the main excitation of the vocal tract, which reduces the biasing effect of the glottal source in the modeling of VTRs.
De-emphasizing the main excitation can also be justified from the observation that this region typically shows large prediction errors that become increasingly dominant with short pitch periods. QCP analysis has previously been shown to be effective in estimating the voice source with GIF \cite{Manu2014}.

\section{Time-varying quasi-closed-phase analysis}
\label{sec:tvqcp}
The spectral estimation and tracking method proposed in this study, TVQCP analysis, combines the ideas of sample selective prediction (i.e., the underlying idea of QCP), sparsity of the prediction error, and long-time nonstationary analysis of the vocal tract system (i.e., the underlying idea of TVLP). In the following, the normal equations of the proposed TVQCP analysis are derived by starting from conventional LP. Note that the optimization schemes in Section~\ref{sec:qcp} all used the $L_2$ norm of the error signal whereas this section uses more general optimization norms. 

\subsection{Linear prediction}
In conventional LP, the current sample $x[n]$ is predicted according to Eq.~(\ref{eq:eq1}) as a linear weighted sum of the past $p$ samples.
By denoting the window size as $N$, the predictor coefficients can be estimated as a solution to the convex optimization problem of generic norm $L_m$ given by
\begin{gather}
\hat{\bm a}=\argmin_a || {\bm x} - {\bm X}{\bm a}||_{m}^m ,\\
\text{where} \quad {\bm x}=[x[0],x[1],\dots,x[N-1]]^T_{_{N\times 1}},\\
{\bm a}=[a_1, a_2,\dots,a_p]^T_{_{p\times 1}},\\
{\bm X}=[X_0,X_1,\dots,X_{N-1}]^T_{_{N\times p}} ,\qquad \text{and}\\
{X_n}=[x[n-1],\dots,x[n-p]]^T_{_{p\times 1}}.
\end{gather}
The minimization of the $L_2$ norm of the residual leads to the least square solution of conventional LP.
However, imposing a sparsity constraint on the residual provides better modeling of both the excitation and vocal tract system.
This is achieved by minimizing the $L_1$ norm of the residual instead of its $L_2$ norm. This change in the optimization norm is known to give a convex approximation of the solution to the $L_0$ norm optimization problem, also referred to as sparse linear prediction (SLP)~\cite{GiacobelloASLP2012,WipfIEEEJSTSP2010}.

\subsection{Weighted linear prediction}
\label{sec:wlpIII}
WLP analysis uses sample-selective prediction and gives differential emphasis to different regions of the speech signal within a glottal cycle (as discussed earlier in Section~\ref{sec:wlp}).
Using a generic $L_m$ norm, WLP can be expressed by minimizing the weighted error signal given by
\begin{gather}
\hat{\bm a}=\argmin_a {\bm W} || {\bm x} - {\bm X}{\bm a}||_{m}^m, 
\end{gather}
where ${\bm W_{N \times N}}$ is a diagonal matrix with its diagonal elements corresponding to a weighting function $w_n$, imposed on the prediction error signal.

\subsection{Time-varying linear prediction}
\label{sec:tvlp}
TVLP is a generalization of conventional LP where the predictor coefficients are continuous functions of time.
Therefore, TVLP can be used in the spectral analysis of nonstationary speech signals using long-time (e.g., 100--200 ms) frames. TVLP imposes a time-continuity constraint on the vocal tract system in the form of low-order basis functions. Due to this time-continuity constraint, TVLP is capable of modeling the slowly varying vocal tract system better than conventional LP that is based on a piecewise constant quasi-stationary approximation.
In TVLP, the current speech sample is predicted using the past $p$ samples as
\begin{gather}
\hat{x}[n]=\sum_{k=1}^{p}a_k[n]x[n-k], 
\end{gather}
where $a_k[n]$ denotes the $k^{th}$ time-varying prediction filter coefficient at time instant $n$.
The time-variant predictor coefficient $a_k[n]$ can be expressed using different basis functions, such as polynomials (i.e., power series), trigonometric series, or Legendre polynomials~\cite{HallSP1983}.
In this study, we use the simple $q^{th}$ order polynomial approximation given by
\begin{gather}
a_k[n]=\sum_{i=0}^{q}{b_{k_i} n^i}.
\end{gather}

The TVLP coefficients are estimated by minimizing the $L_m$ norm of the error signal. This can be presented as the convex optimization problem given by
\begin{gather}
\hat{\bm b}=\argmin_b || {\bm x} - {\bm Y}{\bm b}||_{m}^m , \\
\text{where} \quad {\bm x}=[x[0],x[1],\dots,x[N-1]]^T_{_{N\times 1}} , \\
{\bm b}=[b_{1_0},\dots,b_{1_q},\dots,b_{p_0},\dots,b_{p_q}]^T_{_{p(q+1)\times 1}} , \\
{\bm Y}=[Y_0,Y_1,\dots,Y_{N-1}]^T_{_{N\times p(q+1)}}, \qquad \text{and} \\
{Y_n}=[x[n-1],nx[n-1],\dots,n^q x[n-1],\notag \\
\dots,x[n-p],nx[n-p],\dots, n^q x[n-p]]^T_{_{p(q+1)\times 1}}.
\end{gather}
Again, the $L_2$ and $L_1$ norm minimization lead to the least square solution and the sparse solution to the convex optimization problem respectively~\cite{GiacobelloASLP2012,WipfIEEEJSTSP2010,ChetupalliTVSIcassp2014}. It is to be noted that the $L_2$ norm minimization can be solved in closed form whereas convex optimisation calls for an iterative approach and therefore its computational complexity is larger. The current study uses linear programming in convex optimization for the $L_1$ norm-based methods. Hence, the computational complexity of the $L_1$ norm-based LP methods studied in this article is clearly higher than in the $L_2$ norm-based LP methods.

\subsection{Time-varying weighted linear prediction}
As the final step of the model optimization, let us combine WLP, the technique described in Section~\ref{sec:wlp} and Section~\ref{sec:wlpIII}, and TVLP, the approach presented in Section~\ref{sec:tvlp}. The combination of these two, time-varying weighted linear prediction (TVWLP) analysis, is analogous to WLP where the predictor coefficients are estimated by minimizing the weighted error signal given by
\begin{gather}
\hat{\bm b}=\argmin_b {\bm W} || {\bm x} - {\bm Y}{\bm b}||_{m}^m , \label{eq:eq22}
\end{gather}
where ${\bm W_{N \times N}}$ is a diagonal matrix with its diagonal elements corresponding to the weighting function $w[n]$, imposed on the error signal.

Based on Eq.~(\ref{eq:eq22}), in this study we propose a new TVQCP analysis of speech signals that uses the QCP weighting function (described in Section~\ref{sec:qcpwt}) in matrix ${\bm W}$ of the TVWLP framework above. By using the  $L_1$ norm (i.e., assigning $m=1$ in Eq.~(\ref{eq:eq22})), the TVQCP analysis enables imposing a sparsity constraint on the excitation signal.

\section{Formant tracking experiments}
\label{sec:expt}
One of the main problems with evaluating the performance of a formant tracker and comparing it with other methods is the availability of absolute ground truth in formant frequency values.
It is possible to have such absolute ground truth in the case of synthetic speech signals.
However, there are two limitations with using synthetic speech signals.
The first is that the reference formant frequencies provided by synthetic utterances can be biased towards a particular method of formant tracking if there is a strong similarity in the synthesis model and the analysis model of the tracker.
The second is that the formant trackers are ultimately required to process natural speech signals that do not have any reference ground truth.
The problem with using natural speech signals is the need for a semi-supervised human annotation of the formant frequency value, which by itself can vary from one annotator to another~\cite{lideng2006}.
Formant tracking from natural speech can also be biased by the tools and techniques used for the annotation, such as spectrographic representations and/or methods used for deriving some of the initial estimates.
Also, it should be noted that actual resonance frequencies of the vocal tract cavities need not exactly coincide with the apparent peaks in speech spectra because these spectral peaks might also be harmonics that are a result of the glottal excitation.

In order to address the above problem with reference ground truth, the performance of formant tracking with the proposed TVQCP method is studied using both synthetic and natural speech signals.
Two different types of synthetic signals were used.
In one type, vowels are produced with conventional source-filter modeling of the speech production apparatus using the LF glottal source model and an all-pole vocal tract filter.
In the other type, utterances are generated using physical modeling of the vocal tract and glottal source~\cite{Story2013, Story2016}.
The latter approach is different from the LF source-filter technique because the speech signal is generated based on physical laws, rather than by a digital parametric model similar to the model assumed in LP and it variants.
The physical modeling approach is used to avoid any inherent bias that the LF source-filter technique may have towards the proposed TVQCP method, owing to the fact that both use LP-based methods in vocal tract modeling.

\subsection{Performance metrics}
\label{ssec:metrics}
The formant tracking performance of different methods is evaluated in terms of two different metrics: the formant detection rate (FDR) and formant estimation error (FEE).
Throughout this study, formants are identified by looking for the local peaks of the power spectrum.
The FDR is measured in terms of the percentage of frames where a formant is hypothesized within a specified deviation from the ground truth.
The FDR for the $i^{th}$ formant over $K$ analysis frames is computed as
\begin{align}
D_{i} &=  \frac{1}{K}\sum_{n=1}^{K}{I(\Delta F_{i,n})} ,\\
I(\Delta F_{i,n}) &= \left\{\begin{array}{ll} 1 &\quad\text{if} \left({\Delta F_{i,n}}/{F_{i,n}} < \tau_r \quad \& \quad \Delta F_{i,n} < \tau_a\right) \\ 0 &\quad \text{otherwise, }\end{array}\right. \label{eq:fdr}
\end{align}
where $I(.)$ denotes a binary formant detector function and $\Delta{F_{i,n}}=|F_{i,n}-\hat{F}_{i,n}|$ is the absolute deviation of the hypothesized formant frequency $\hat{F}_{i,n}$ for the $i^{th}$ formant at the $n^{th}$ frame from the reference ground truth $F_{i,n}$.
The thresholds $\tau_r$ and $\tau_a$ denote the relative deviation and absolute deviation respectively.

Using a single detection threshold, either a relative threshold or an absolute threshold, is problematic on a linear frequency scale.
For higher formants, the relative deviation needs to be smaller than that for the lower formants.
Similarly, the absolute deviation for lower formants needs to be smaller than that for the higher formants.
In order to define a common detection strategy for all formants, two thresholds, one on relative deviation and the other on absolute deviation, must be used. The relative threshold controls the detection rates of lower formants whereas the absolute threshold controls the detection rates of higher formants.

The FEE is measured in terms of the average absolute deviation of the hypothesized formants from the ground truth.
The FEE for the $i^{th}$ formant over $K$ analysis frames is computed as
\begin{equation}
R_i=\frac{1}{K}\sum_{n=1}^{K}{\Delta F_{i,n}}.
\end{equation}
The FDR and FEE values are only computed for frames that are voiced or for some particular phonetic category of interest.
One problem with accumulating FEEs over all frames is that a few large error outliers can dictate the overall score.
This is even more severe for the root mean square error (RMSE) criterion that is a widely used metric for measuring formant estimation accuracy.
In view of this, we propose using mean absolute error, which is less sensitive to outliers, as a measure for FEE.
The reading of FEE scores in conjunction with FDR scores, which denote the number of frames detected within a fixed threshold, can give a better sense of the performance of a formant tracker.

\vspace*{-0.2cm}
\subsection{The choice of window size and polynomial order}
As outlined in Section~\ref{sec:tvlp}, TVLP analysis involves two parameters (in addition to prediction order $p$) that need to be set: window size $N$ and polynomial order $q$.
Longer window sizes (e.g., 500 ms) are useful for the efficient parameterization of speech signals but would introduce longer delays.
Moreover, longer window sizes require higher polynomial orders in order to model the time-varying characteristics of the vocal tract and can lead to computational problems due to the inversion of rank deficient matrices. Therefore, moderate window sizes (e.g., 100--200 ms) are a good overall compromise that enables the efficient parameterization of the slowly time-varying characteristics of the vocal tract using low-order polynomials (e.g., $q=3$). 
\begin{figure}[t]
\vspace*{-0.2cm}
\includegraphics[width=\columnwidth,trim=1.2cm 0 1.5cm 0,clip]{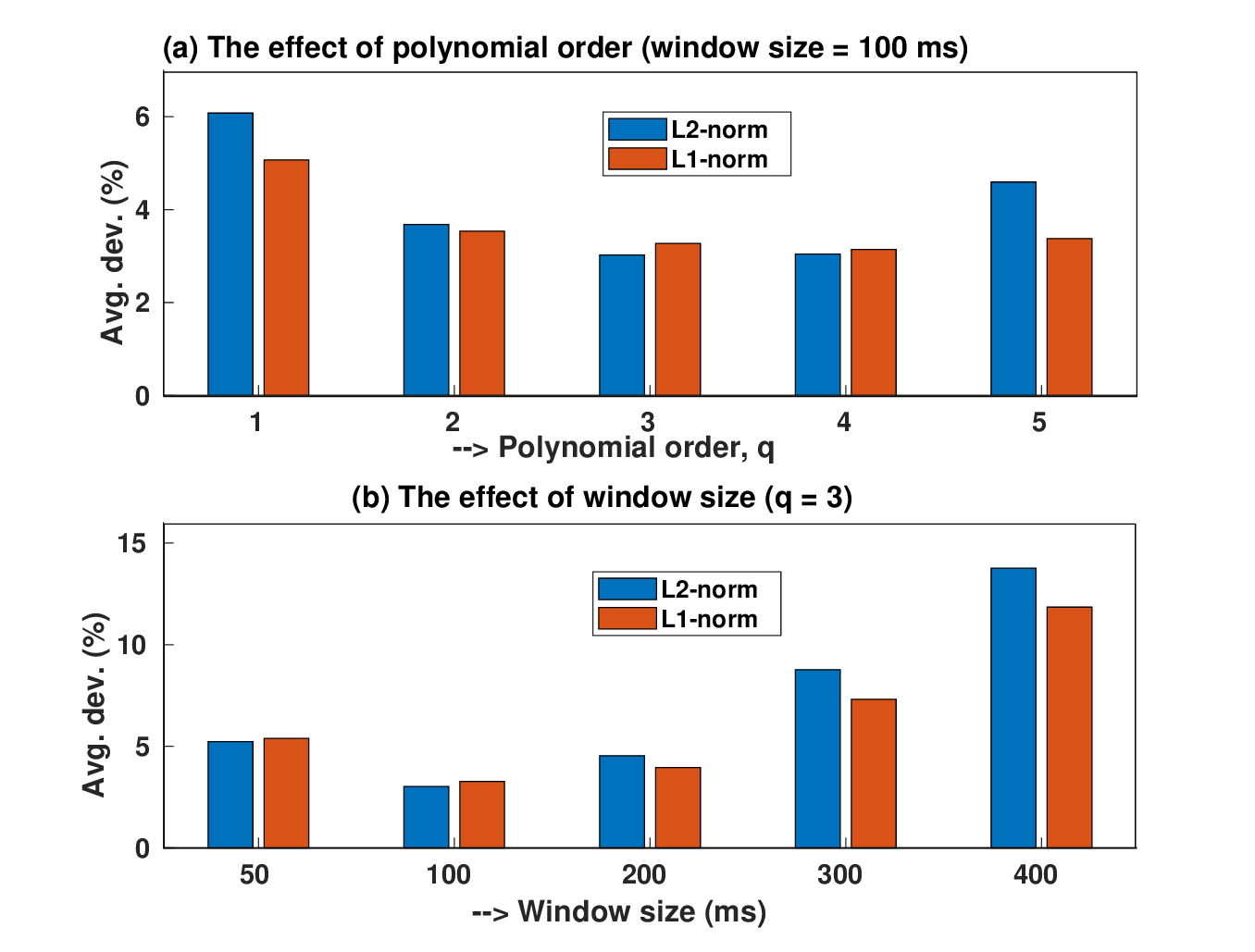}
\vspace*{-0.6cm}
\caption{\label{fig:fig2} Relative deviation (in percentage) of the TVLP-estimated formants from their ground truth, averaged over the three first formants as a function of (a) polynomial order and (b) window size.}
\vspace*{-0.5cm}
\end{figure}

In order to study the choice of the window size and polynomial order in TVLP analysis, an initial experiment was conducted on a set of synthetic utterances. The effect of these two parameters on a larger dataset of natural speech utterances will be studied later.
The synthetic speech utterances were generated starting with ten (5 male, 5 female) randomly chosen natural speech utterances from the TIMIT-VTR database~\cite{lideng2006}.
The natural utterance was first inverse filtered using a high order ($p=18$) short-time LP analysis (20-ms frame size, 10-ms frame shift, and a sampling rate of 8 kHz) to compute a spectrally flat residual signal that was void of any formant structure.
This residual signal was then used to excite an $8^{th}$ order all pole model constructed using the first four reference formants and bandwidths available for the utterances as part of the VTR database~\cite{lideng2006}.

\begin{figure}[h]
\vspace*{-0.2cm}
\includegraphics[width=\columnwidth, height=7cm,trim=0.6cm 0 1cm 0,clip]{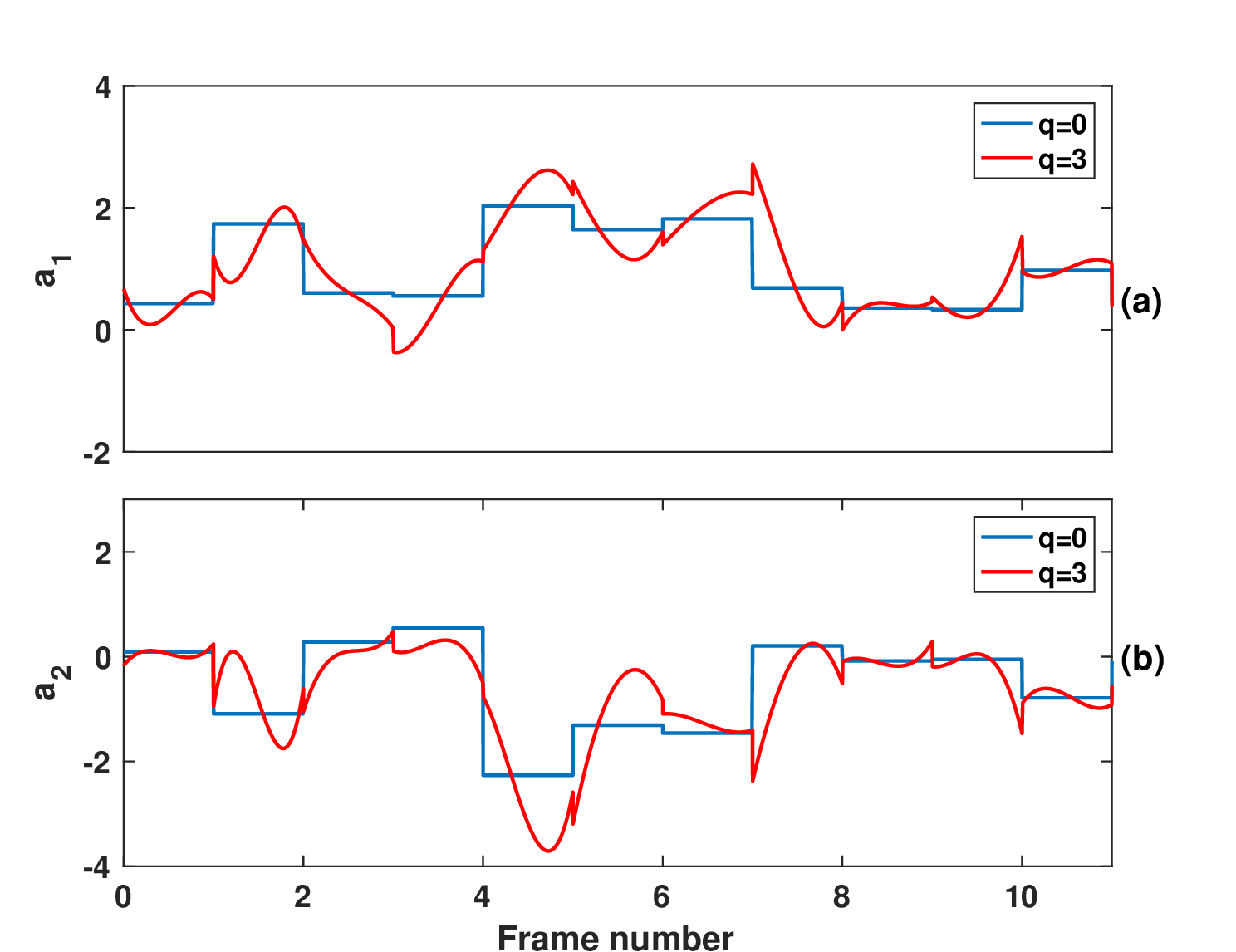}
\caption{\label{fig:akplot} The trajectories of $a_k[n]$ for q=0 and q=3 in TVLP-L2 for the first and second coefficients ($a_1$ and $a_2$)  are shown in (a) and (b), respectively. The word $'materials'$ produced by a female talker is used for the illustration.}
\vspace*{-0.2cm}
\end{figure}

The results of the experiment are shown in Fig.~\ref{fig:fig2}, which depicts the relative deviation of the estimated formants from their ground truth, averaged over the first three formants ($F_1$, $F_2$, and $F_3$) for different values of polynomial order and window size. TVLP analyses computed using the $L_2$ norm are shown as blue bars and those computed with the $L_1$ norm are shown as red bars. Fig.~\ref{fig:fig2}(a) depicts the TVLP performance using  a fixed window size of 100 ms but with the varying polynomial order $q$. It can be seen that the best performance is obtained for polynomial orders between $q=2$ and $q=4$ and the performance starts to deteriorate at the order of $q=5$. Similarly, Fig.~\ref{fig:fig2}(b) shows the performance by varying the window size at a fixed polynomial order of $q=3$.
It can be seen that the performance is good with moderate window sizes of 100 ms and 200 ms, but the performance starts to deteriorate for longer window sizes. Therefore, in the experiments that follow in the remainder of the paper, we used a window size of 100 ms and a polynomial order of $q=3$ in time-varying LP analyses. An example of using two different polynomial orders ($q=0$ and $q=3$) for an utterance produced by a female speaker is shown in Fig.~\ref{fig:akplot}. The figure depicts the contours of the two lowest coefficients ($a_1$ and $a_2$) computed using TVLP with the $L_2$ norm. It can be seen that the filter taps computed using $q=0$ and $q=3$ follow a similar general trend over the entire time-span shown in the figure but the contours computed using $q=3$ are clearly more dynamic and their values change also during each frame.

\begin{figure}[b]
\vspace*{-0.2cm}
\includegraphics[width=\columnwidth,trim=0cm 0 0cm 0,clip]{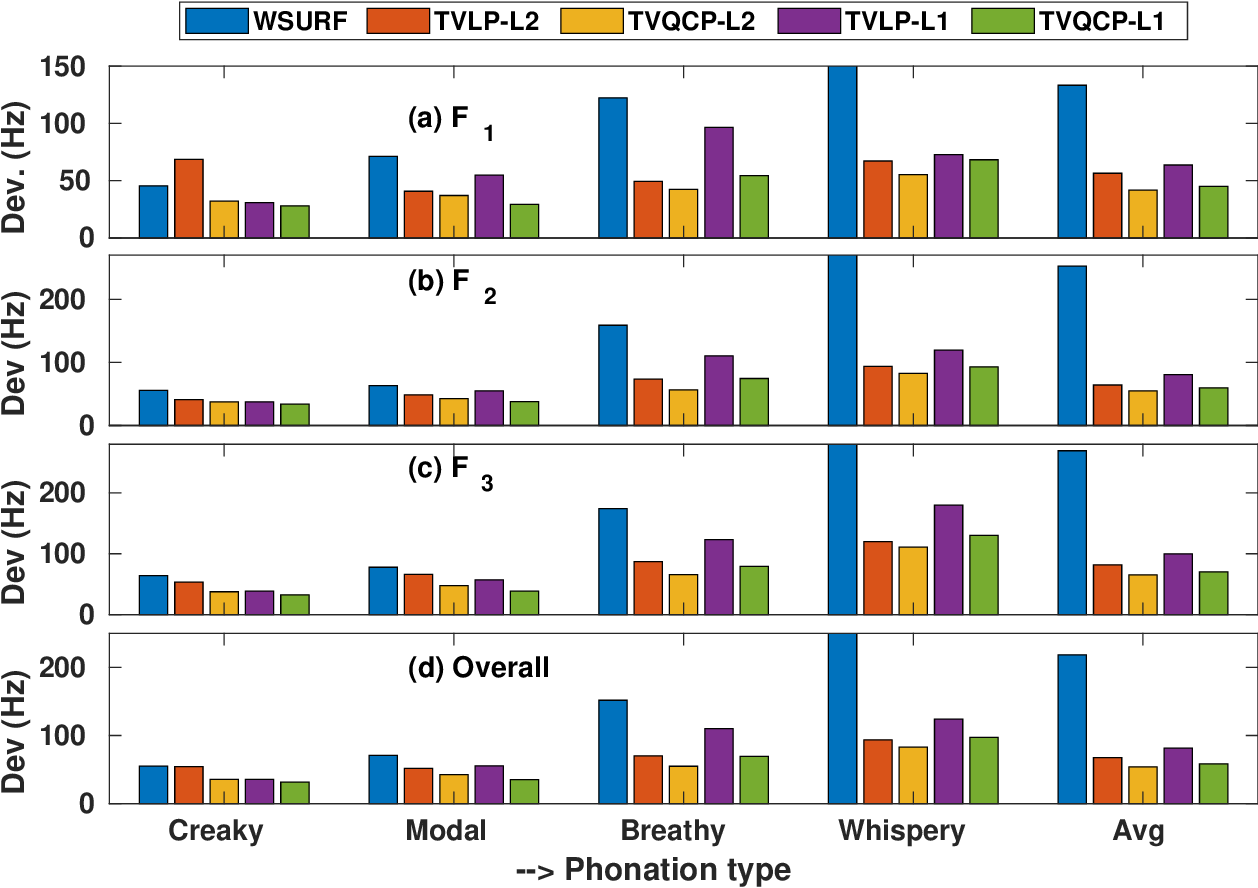}
\caption{\label{fig:fig3} The absolute deviation (FEE) of the estimated first three formants ($F_1$, $F_2$, and $F_3$) from their ground truth and their overall average for different phonation types of the LF model--based synthetic data.}
\vspace*{-0.2cm}
\end{figure}

\begin{figure}[b]
\vspace*{-0.2cm}
\includegraphics[width=\columnwidth,trim=0cm 0 0cm 0,clip]{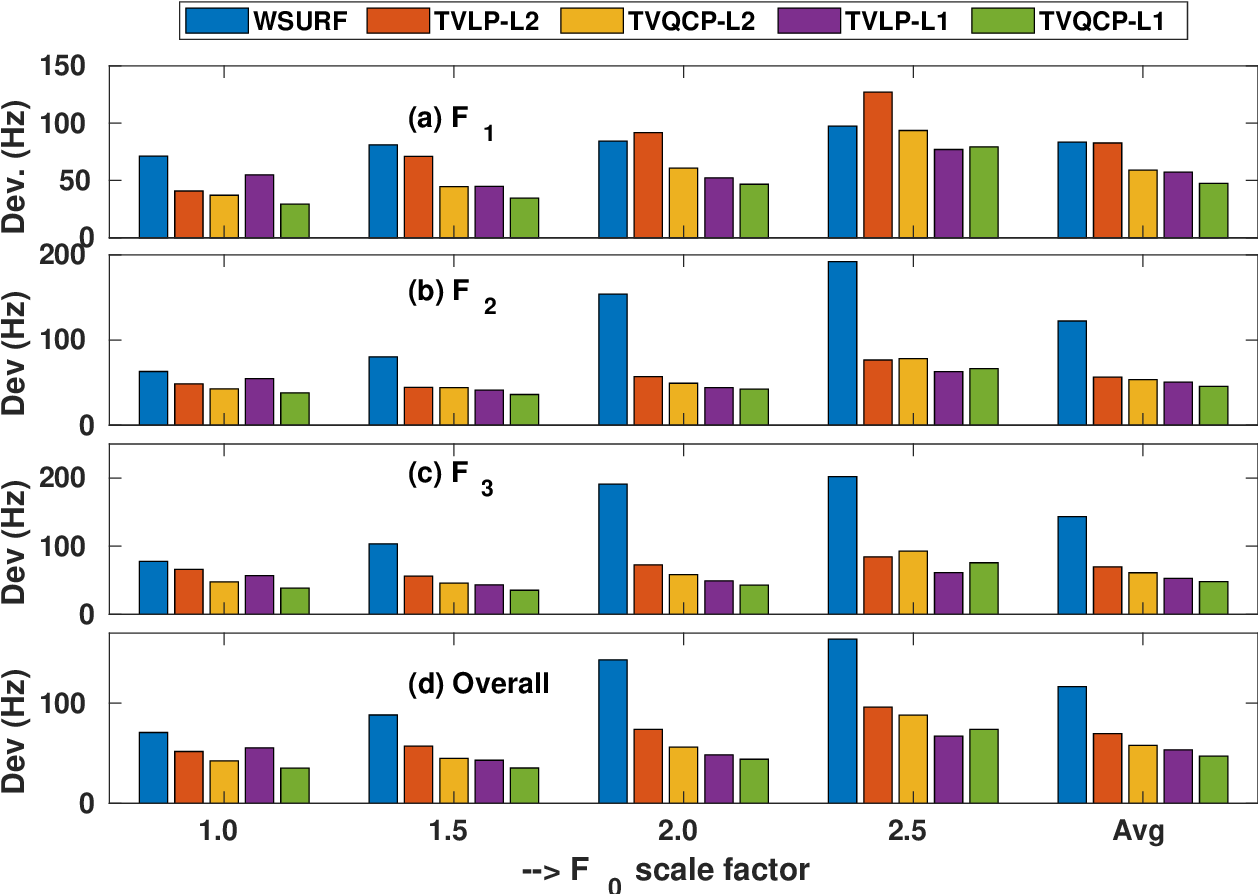}
\caption{\label{fig:fig4} The absolute deviation (FEE) of the estimated first three formants ($F_1$, $F_2$, and $F_3$) from their ground truth and their overall average for different mean $F_0$ values of the LF model--based synthetic data.}
\vspace*{-0.2cm}
\end{figure}

\subsection{Experiments on LF model--based synthetic data}
The performance of the proposed TVQCP method in formant tracking is studied next in this section by analyzing how the method's performance is affected by variations in the glottal excitation (both in fundamental frequency and phonation type). Formant tracking provided by the TVQCP method is compared with that of TVLP using both the $L_1$ norm and $L_2$ norm. In addition, a comparison with the traditional LP covariance--based method (known as the entropic signal processing system (ESPS) method~\cite{talkin1987}) used in the popular open source tool Wavesurfer~\cite{wavesurfer2000} (denoted by ``WSURF") is also provided.

The TVQCP and TVLP analyses are carried out over non-overlapping 100-ms windows using a prediction order of $p=8$ and a polynomial order of $q=3$.
The ESPS method used in Wavesurfer adopts a short-time (25-ms Hamming window, 10-ms frame shift) $12^{th}$ order stabilized covariance--based LP analysis followed by a dynamic programming--based tracking of formants~\cite{wavesurfer2000}.

\subsubsection{The dataset}
Four different phonation types (creaky, modal, breathy, and whispery phonation) and four different ranges of fundamental frequency (mean utterance $F_0$ scaled by the factors 1.0, 1.5, 2.0, and 2.5) are considered for generating the synthetic speech test utterances.
The phonation type and $F_0$ range are controlled by using the LF model for the glottal source~\cite{fant1985}.
The LF source parameter values used to synthesize the different phonation types in the current study are taken from \cite{Gobl2003,gobl1989preliminary}.

The four different fundamental frequency ranges are generated by scaling the original $F_0$ contour of a natural speech utterance (3--5 sec long) by different factors before synthesizing the speech signal.
A modal LF excitation is generated based on the new $F_0$ contour while retaining the original rate of formants and hence keeping the speaking rate intact.
Speech signals are synthesized by filtering the LF glottal flow derivative signal using an all-pole model with the first four semi-automatically derived formants and bandwidths of the natural utterance part of the VTR database~\cite{lideng2006}.
Ten randomly selected utterances (5 male and 5 female) from the VTR database are synthesized for the four different phonation types and four different mean $F_0$ at a sampling rate of 8 kHz.

\subsubsection{The effect of phonation type}
The performance of the TVQCP and TVLP methods are shown in Fig.~\ref{fig:fig3} for the four different phonation types.
The TVQCP method that minimizes the $L_2$ norm (denoted by TVQCP-L2) performed best overall, marginally better than the TVQCP method that minimizes the $L_1$ norm (denoted by TVQCP-L1). The $L_1$ norm minimization seemed to perform better than the $L_2$ norm for most cases in creaky and modal phonations while the $L_2$ norm performed better for breathy and whispery phonations that exhibit larger open quotients and higher spectral tilts.
Overall, it can be seen that the TVQCP methods performed better than their TVLP counterparts across all formants and all phonation types. Moreover, the performance of the both TVLP and TVQCP methods is clearly better than that of the popular Wavesurfer tool.

\subsubsection{The effect of fundamental frequency}
The performance of the TVQCP and TVLP methods are shown in Fig.~\ref{fig:fig4} for all the four ranges of $F_0$ values.
It can be seen that TVQCP optimized using both the $L_1$ and $L_2$ norms provided consistent improvements over TVLP up to a scale factor of 2.0.
The mixed performance for the scale factor 2.5 may be due to the new $F_0$ values moving very close to $F_1$ in the synthetic utterances.
However, it has been observed that this minor aberration gets corrected if $F_1$ is shifted upward by a small percentage.
Also, the $L_1$ norm optimization seemed to perform better than the $L_2$ norm in most cases except for TVLP for $F_1$ and $F_2$ at $F_0$ scale factor 1.0. In terms of overall performance across all fundamental frequency ranges and formants, TVLP-L2, TVQCP-L2, TVLP-L1, and TVQCP-L1 showed a consistent improvement in this order.

The overall performance of the formant tracking methods is given in Table~\ref{tab:tab1} by averaging over all phonation types and $F_0$ ranges.
\begin{table}
\centering
\caption{\label{tab:tab1} The absolute deviation (FEE in Hz) of the estimated first three formants ($F_1$, $F_2$, and $F_3$) from their ground truth and their overall average over all phonation types and fundamental frequencies of the LF model--based synthetic data.}
\vspace{-0.2cm}
\setlength{\tabcolsep}{2pt}
{\small
\begin{tabular}{|c||c|c|c|c|c|}
\hline
& WSURF & TVLP-L2 & TVQCP-L2 & TVLP-L1 & TVQCP-L1 \\\hline\hline
\multicolumn{6}{|l|}{{\bf (a) Avg. over all phonation types}} \\\hline
$F_1$ & 133.3 &  56.5 &  {\bf 41.8} &  63.7 &  45.0 \\\hline
$F_2$ & 252.7 &  64.2 &  {\bf 54.8} &  80.6 &  59.8 \\\hline
$F_3$ & 269.3 &  81.6 &  {\bf 65.4} &  99.7 &  70.1 \\\hline\hline
Avg & 218.4 &  67.4 &  {\bf 54.0} &  81.3 &  58.3 \\\hline \hline
\multicolumn{6}{|l|}{{\bf (b) Avg. over all $F_0$ range}} \\\hline
$F_1$ & 83.5 &  82.7 &  59.0 &  57.2 &  {\bf 47.5} \\\hline
$F_2$ & 122.4 &  56.6 &  53.6 &  50.7 &  {\bf 45.7} \\\hline
$F_3$ & 143.5 &  69.7 &  61.0 &  52.6 &  {\bf 48.1} \\\hline\hline
Avg & 116.5 &  69.7 &  57.9 &  53.5 &  {\bf 47.1} \\\hline\hline
\multicolumn{6}{|l|}{{\bf (c) Avg. over all phonation types and $F_0$ range}} \\\hline
Overall & 167.5 & 68.6 & 56.0 & 67.4 & {\bf 52.7} \\\hline
\end{tabular}
}
\vspace*{-0.5cm}
\end{table}
The general observation is that the FEE reduced considerably with the use of QCP analysis (TVQCP analysis vs. TVLP) and that there is a marginal reduction when using the sparsity constraint ($L_1$ norm vs. $L_2$ norm).
Overall, both the TVLP and TVQCP methods provided large improvements over the popular Wavesurfer tool with 60 to 70 percentage unit reduction in the estimation error.

\subsection{Experiments on simulated high-pitched child speech using a physical modeling approach}
The formant estimation accuracy of the proposed TVQCP method is compared to that of TVLP using synthetic data generated by an alternate, physical modeling approach of the speech production apparatus~\cite{Story2013}.
The experiments in this section try to address two issues with the evaluation of formant estimation and tracking methods.
One is the bias of the LF model--based synthetic data towards LP--based methods, and the other is the performance of these methods on speech signals at very high fundamental frequencies.

An $8^{th}$ order analysis is used for all the methods, and the original data at 44.1 kHz is downsampled to 16 kHz and passed through a preemphasis filter $P(z)=1-0.97z^{-1}$ before further processing.
The TVLP and TVQCP methods use a 100-ms window size and a polynomial order of $q=3$.
The final formant estimates are evaluated at a 20-ms frame shift to match the reference formants rate.

\subsubsection{The dataset}
The simulated data consists of eight short child speech utterances of a high pitch (as high as 600 Hz) used in \cite{Story2016}.
The eight utterances include two steady vowels, [a] and [i], of  340 ms duration each with a constant $F_0$ of 400 Hz.
The six simulations of 1.03 s each are three time-varying vocal tract shapes combined with two different time-dependent $F_0$ variations.
The three time-varying vocal tract shapes correspond to the sequence of sounds \{i.a.i.a.i.a\}, \{ae.u.ae.u.ae.u\}, and \{i.a.i\}.
The fundamental frequency of the utterances varies between 240 Hz to 500 Hz, one in a smooth increasing--decreasing pattern and the other in a reverse pattern over the entire length of the utterance.
All the utterances have four vocal tract resonances and are stored at a 44.1 kHz sampling rate.
More information on the formant and $F_0$ contours used and other details of the dataset can be found in \cite{Story2016}.

\subsubsection{The results}
FEEs computed using both the $L_1$ and $L_2$ norms in the TVLP and TVQCP methods are given in Table~\ref{tab:pmodel}. It is seen that the TVQCP method tends to give a consistent shift in estimating the fourth formant.
This could be due to many reasons including the pre-emphasis, sampling rate, model limitations, limited synthetic data, and this needs further investigation.
In view of this, further discussions in this section are limited to the first three formants.
\begin{table}
\centering
\caption{\label{tab:pmodel} The absolute deviation (FEE in Hz) of the estimated first four formants ($F_1$--$F_4$) from their ground truth on child speech generated using the physical modeling approach.}
\vspace{-0.2cm}
\begin{tabular}{|c||c|c|c|c|}
\hline
Method & $F_1$ & $F_2$ & $F_3$ & $F_4$ \\\hline\hline
TVLP-L2 &  70.8 & 163.9 &  69.3 &  {\bf76.8} \\\hline
TVLP-L1 &  52.8 & 105.0 &  61.4 & 106.4 \\\hline
TVQCP-L2&  {\bf32.9} &  51.9 &  61.4 & 136.1 \\\hline
TVQCP-L1&  33.0 &  {\bf 48.3} &  {\bf 54.4} & 148.4 \\\hline
\end{tabular}
\vspace*{-0.5cm}
\end{table}
It can be seen from the table that imposing a sparsity constraint with the $L_1$ norm minimization clearly improves the accuracy of TVLP and TVQCP.
The continuity constraint imposed by time-varying models (TVLP) do not seem to provide much improvement on their own. However, when combined with the QCP weighting, the continuity constraint seems to provide large improvements in the case of TVQCP-L1 and some marginal improvement in the case of TVQCP-L2.
Owing to the limited availability of data, 
it may not be possible to draw too many inferences from this experiment.
Nevertheless, it demonstrates the usefulness of combining the ideas of QCP analysis, time-varying linear predictive analysis, and the sparsity constraint for formant tracking applications.

\subsection{Experiments on natural speech data}
One of the primary goals of this paper is to evaluate the performance of the proposed TVQCP--based formant tracker on real speech utterances.
A detailed evaluation of the TVQCP method and a comparison with some of the state-of-the-art formant trackers is presented in this section.

\subsubsection{The dataset}
The performance of different methods in formant tracking was evaluated on natural speech signals using the VTR database published in~\cite{lideng2006}.
The test data of the VTR database is used for the evaluation and this data consists of 192 utterances (8 utterances pronounced by 8 female and 16 male speakers). The duration of each utterance varies between 2 and 5 s.
The first four reference formant frequency and bandwidth values derived using a semi-supervised LP--based algorithm \cite{lideng2004} are provided for every 10-ms interval.
The first three reference formant frequency values have been verified and corrected manually based on spectrographic evidence.
All the speech data, originally recorded at a 16 kHz sampling rate, are downsampled to 8 kHz before processing.
A pre-emphasis filter of $P(z)=1-0.97z^{-1}$ is used to preprocess the speech signals.
Based on our earlier experiments on formant tracking using synthetic speech signals, we use a default window size of 100 ms, a prediction order of 8, and a polynomial order of 3 for the time-varying linear predictive methods unless otherwise mentioned.
All the performance metrics presented in this section are average scores computed over vowels, diphthongs, and semivowels. These are phonetic categories whose manually corrected formant ground truths are more reliable compared to other categories.



\begin{table}[t]
\centering
\caption{\label{tab:win_lpo_poly1} The effect of window size (in ms), prediction order, and polynomial order on the formant tracking performance of TVQCP-L1.}
\vspace{-0.2cm}
{\small
\begin{tabular}{|c|c|c|c||c|c|c|}
\hline
 & \multicolumn{3}{c||}{FDR (\%)} & \multicolumn{3}{|c|}{FEE (Hz)} \\\cline{2-7}
 & ~$F_1$~ & ~$F_2$~ & ~$F_3$~ & ~$\delta F_1$~ & ~$\delta F_2$~ & ~$\delta F_3$\\\hline
\multicolumn{7}{c}{}\\\hline
$N_t$ & \multicolumn{6}{c|}{Effect of window size ($p$=8, $q$=3)} \\\hline
50 ms  & 88.8 & 94.9 & 90.1 & 79.2 & 102.0 & 138.0 \\\hline
100 ms & {\bf90.9} & {\bf96.5} & 92.2 & 67.6 & {\bf91.7} & 123.9  \\\hline
200 ms & 90.8 & {\bf96.5} & {\bf92.6} & {\bf66.6} & 93.6 & {\bf123.0}  \\\hline
\multicolumn{7}{c}{}\\\hline
$p$ & \multicolumn{6}{c|}{Effect of prediction order ($N_t$=100 ms, $q$=3)}\\\hline
6 & 65.9 & 65.3 & 21.7 & 224.9 & 454.8 & 919.9 \\\hline
7 & 75.9 & 81.5 & 54.6 & 151.0 & 238.8 & 459.4 \\\hline
8 & 90.9 & {\bf96.5} & {\bf92.2} & 67.6 & {\bf91.7} & {\bf123.9}  \\\hline
9 & {\bf92.0} & 89.9 & 84.1 & {\bf63.0} & 159.0 & 198.4  \\\hline
10 & 91.9 & 71.6 & 57.6 & 63.6 & 309.2 & 433.2 \\\hline
\multicolumn{7}{c}{}\\\hline
$q$ & \multicolumn{6}{c|}{Effect of polynomial order ($N_t$=100 ms, $p$=8)}\\\hline
0 & 88.3 & 92.5 & 89.4 & 72.2 & 114.4 & 143.3  \\\hline
1 & {\bf91.0} & 96.5 & 92.8 & {\bf65.4} & 90.3 & {\bf119.5}  \\\hline
2 & {\bf91.0} & {\bf96.6} & {\bf92.5} & 66.3 & {\bf90.2} & 121.4 \\\hline
3 & 90.9 & 96.5 & 92.2 & 67.6 & 91.7 & 123.9  \\\hline
\end{tabular}
}
\vspace*{-0.1cm}
\end{table}

\begin{table}[t]
\centering
\caption{\label{tab:wt_fn1} The effect of different weighting functions on the formant tracking performance of TVWLP ($L_1$ norm).}
\vspace{-0.2cm}
{\small
\begin{tabular}{|c|c|c|c||c|c|c|}
\hline
 Weighting & \multicolumn{3}{c||}{FDR (\%)} & \multicolumn{3}{|c|}{FEE (Hz)} \\\cline{2-7}
 func. & ~$F_1$~ & ~$F_2$~ & ~$F_3$~ & ~$\delta F_1$~ & ~$\delta F_2$~ & ~$\delta F_3$\\\hline\hline
STE      & 89.1 & 95.1 & 90.5 & 73.6 & 105.1 & 141.3 \\\hline
Residual & 90.6 & 96.3 & 91.8 & 68.2 & 93.2 & 126.9  \\\hline
QCP      & {\bf90.9} & {\bf96.5} & {\bf92.2} & {\bf67.6} & {\bf91.7} & {\bf123.9} \\\hline
\end{tabular}
}
\vspace*{-0.3cm}
\end{table}

\begin{table}[t]
\centering
\caption{\label{tab:gci1} The effect of GCI detection errors on formant tracking performance with TVQCP-L1. $Rerr$ and $Ferr$ refer to random and fixed error, respectively. The effect of the duration quotient (DQ) of the QCP weighting function is also presented.}
\vspace{-0.2cm}
{\small
\begin{tabular}{|c|c|c|c||c|c|c|}
\hline
 & \multicolumn{3}{c||}{FDR (\%)} & \multicolumn{3}{|c|}{FEE (Hz)} \\\cline{2-7}
 & ~$F_1$~ & ~$F_2$~ & ~$F_3$~ & ~$\delta F_1$~ & ~$\delta F_2$~ & ~$\delta F_3$\\\hline
\multicolumn{7}{c}{}\\\hline
$Rerr$ & \multicolumn{6}{c|}{The effect of GCI error}\\\hline
0 & {\bf90.9} & {\bf96.5} & {\bf92.2} & {\bf67.6} & {\bf91.7} & {\bf123.9}  \\\hline
4 & {\bf90.9} & 96.3 & 92.0 & 68.4 & 92.7 & 125.3  \\\hline
8 & 90.7 & 96.1 & 91.7 & 69.1 & 94.6 & 127.6  \\\hline
12& 90.6 & 96.2 & 91.9 & 69.1 & 94.1 & 127.7  \\\hline
16& 90.6 & 96.0 & 91.5 & 69.2 & 94.9 & 129.9  \\\hline
\multicolumn{7}{c}{}\\\hline
$Ferr$ & \multicolumn{6}{c|}{The effect of GCI error}\\\hline
-16 & 90.1 & 95.8 & 91.4 & 71.0 & 98.7 & 133.8 \\\hline
-8  & 90.5 & 96.0 & 91.3 & 69.3 & 96.1 & 132.1  \\\hline
-4  & 90.3 & 96.0 & 91.4 & 68.8 & 95.1 & 131.2 \\\hline
0 & {\bf90.9} & {\bf96.5} & {\bf92.2} & {\bf67.6} & {\bf91.7} & {\bf123.9} \\\hline
4 & 90.8 & 96.1 & 91.9 & 71.0 & 95.9 & 125.5  \\\hline
8 & 90.5 & 95.9 & 91.3 & 72.2 & 97.7 & 130.9  \\\hline
16 & 88.5 & 95.6 & 90.1 & 74.4 & 100.4 & 143.2  \\\hline 
\multicolumn{7}{c}{}\\ \hline
$DQ$ & \multicolumn{6}{c|}{The effect of DQ}\\\hline
0.5 & 90.8 & 96.0 & 91.7 & 70.7 & 97.0 & 129.1  \\\hline
0.6 & 90.8 & 96.0 & 92.0 & 69.5 & 95.9 & 127.1  \\\hline
0.7 & {\bf90.9} & 96.3 & 92.0 & 68.1 & 93.6 & 125.2  \\\hline
0.8 &{\bf90.9} & {\bf96.5} & {\bf92.2} & {\bf67.6} & {\bf91.7} & {\bf123.9} \\\hline
0.9 & {\bf90.9} & 96.3 & {\bf92.2} & 68.0 & 92.4 & 124.5  \\\hline
\end{tabular}
}
\vspace*{-0.1cm}
\end{table}

\begin{table}[t]
\centering
\caption{\label{tab:win_lpo_poly} The effect of window size (in ms), prediction order, and polynomial order on the formant tracking performance of TVQCP-L2.}
\vspace{-0.3cm}
{\small
\begin{tabular}{|c|c|c|c||c|c|c|}
\hline
 & \multicolumn{3}{c||}{FDR (\%)} & \multicolumn{3}{|c|}{FEE (Hz)} \\\cline{2-7}
 & ~$F_1$~ & ~$F_2$~ & ~$F_3$~ & ~$\delta F_1$~ & ~$\delta F_2$~ & ~$\delta F_3$\\\hline
\multicolumn{7}{c}{}\\\hline
$N_t$ & \multicolumn{6}{c|}{Effect of window size ($p$=8, $q$=3)} \\\hline
50 ms & {\bf 90.9} & 94.4 & 90.2 & 67.8 & 111.0 & 139.7 \\\hline
100 ms &  90.6 & {\bf 96.1} & {\bf 92.0} &  68.4 & {\bf 94.5} & {\bf 126.2} \\\hline
200 ms & 89.9 & 93.9 & 91.0 & {\bf68.3} & 118.7 & 137.9 \\\hline
\multicolumn{7}{c}{}\\\hline
$p$ & \multicolumn{6}{c|}{Effect of prediction order ($N_t$=100 ms, $q$=3)}\\\hline
6 & 66.8 & 65.6 & 21.9 & 226.9 & 454.2 & 922.1 \\\hline
7 & 75.5 & 78.6 & 53.0 & 152.7 & 261.8 & 477.0 \\\hline
8 & 90.6 & {\bf 96.1} & {\bf 92.0} & 68.4 & {\bf 94.5} & {\bf 126.2} \\\hline
9 & 91.6 & 89.0 & 83.6 & 62.5 & 165.3 & 201.6 \\\hline
10 & {\bf 92.2} & 72.1 & 55.2 & {\bf 61.3} & 292.6 & 448.9 \\\hline
\multicolumn{7}{c}{}\\\hline
$q$ & \multicolumn{6}{c|}{Effect of polynomial order ($N_t$=100 ms, $p$=8)}\\\hline
0 & 87.8 & 90.6 & 87.3 & 73.1 & 125.4 & 157.8 \\\hline
1 & 90.6 & 94.5 & 91.4 & {\bf 65.7} & 112.6 & 132.8 \\\hline
2 & {\bf 90.7} & 94.3 & 91.2 & 66.5 & 113.3 & 135.5 \\\hline
3 & 90.6 & {\bf 96.1} & {\bf 92.0} & 68.4 & {\bf 94.5} & {\bf 126.2} \\\hline
\end{tabular}
}
\vspace*{-0.2cm}
\end{table}

\begin{table}[t]
\centering
\caption{\label{tab:wt_fn} The effect of different weighting functions on the formant tracking performance of TVWLP ($L_2$ norm).}
\vspace{-0.2cm}
{\small
\begin{tabular}{|c|c|c|c||c|c|c|}
\hline
 Weighting & \multicolumn{3}{c||}{FDR (\%)} & \multicolumn{3}{|c|}{FEE (Hz)} \\\cline{2-7}
 func. & ~$F_1$~ & ~$F_2$~ & ~$F_3$~ & ~$\delta F_1$~ & ~$\delta F_2$~ & ~$\delta F_3$\\\hline\hline
STE & 87.4 & 93.0 & 88.1 & 80.0 & 121.9 & 161.2 \\\hline
Residual & 89.6 & 95.9 & 91.6 & {\bf 68.3} & 96.8 & 132.1 \\\hline
QCP & {\bf 90.6} & {\bf 96.1} & {\bf 92.0} &  68.4 & {\bf 94.5} & {\bf 126.2} \\\hline
\end{tabular}
}
\vspace*{-0.2cm}
\end{table}


\begin{table}[t]
\centering
\caption{\label{tab:gci} The effect of GCI detection errors on formant tracking performance with TVQCP-L2. $Rerr$ and $Ferr$ refer to random and fixed error, respectively. The effect of the duration quotient (DQ) of the QCP weighting function is also presented.}
\vspace{-0.2cm}
{\small
\begin{tabular}{|c|c|c|c||c|c|c|}
\hline
 & \multicolumn{3}{c||}{FDR (\%)} & \multicolumn{3}{|c|}{FEE (Hz)} \\\cline{2-7}
 & ~$F_1$~ & ~$F_2$~ & ~$F_3$~ & ~$\delta F_1$~ & ~$\delta F_2$~ & ~$\delta F_3$\\\hline
\multicolumn{7}{c}{}\\\hline
$Rerr$ & \multicolumn{6}{c|}{The effect of GCI error}\\\hline
0 & {\bf 90.6} & {\bf 96.1} & {\bf 92.0} & {\bf 68.4} & {\bf 94.5} & {\bf 126.2} \\\hline
4 & 90.3 & 96.0 & 91.4 & 68.9 & 95.5 & 130.2 \\\hline
8 & 89.9 & 95.5 & 91.0 & 69.7 & 99.3 & 134.8 \\\hline
12 & 89.6 & 95.5 & 90.9 & 69.8 & 98.3 & 136.7 \\\hline
16 & 89.3 & 95.4 & 90.7 & 70.6 & 99.8 & 138.0 \\\hline
\multicolumn{7}{c}{}\\\hline
$Ferr$ & \multicolumn{6}{c|}{The effect of GCI error}\\\hline
-16 & 87.7 & 94.4 & 89.1 & 75.5 & 109.8 & 150.8 \\\hline
-8 & 88.5 & 95.1 & 90.0 & 72.0 & 103.2 & 143.9 \\\hline
-4 & 89.0 & 95.3 & 90.5 & 71.1 & 100.9 & 141.1 \\\hline
0 & 90.6 & {\bf 96.1} & {\bf 92.0} & {\bf 68.4} & {\bf 94.5} & {\bf 126.2} \\\hline
4 & {\bf 90.7} & 95.7 & 91.4 & 71.8 & 97.8 & 128.6 \\\hline
8 & 89.8 & 95.6 & 91.0 & 73.6 & 99.8 & 132.1 \\\hline
16 & 87.0 & 92.6 & 85.5 & 80.9 & 121.4 & 171.3 \\\hline
\multicolumn{7}{c}{}\\\hline
$DQ$ & \multicolumn{6}{c|}{The effect of DQ}\\\hline
0.5 & 90.6 & 95.9 & 91.8 & 69.8 & 96.6 & 127.7 \\\hline
0.6 & {\bf 90.8} & 96.1 & 92.0 & 68.9 & 95.0 & 126.1 \\\hline
0.7 & 90.6 & 96.1 & 92.0 & 68.4 & 94.5 & 126.2 \\\hline
0.8 & 90.6 & {\bf 96.2} & 92.0 & {\bf 68.2} & {\bf 94.0} & {\bf 125.9}  \\\hline
0.9 & 90.5 & 96.1 & {\bf 92.1} & 68.5 & 94.5 & 126.5 \\\hline
\end{tabular}
}
\vspace*{-0.2cm}
\end{table}

\subsubsection{The effect of window size, prediction order, and polynomial order}
The effect of the choices for the window size, prediction order, and polynomial order for the tracking performance of the TVQCP-L1 and TVQCP-L2 methods is provided in Table~\ref{tab:win_lpo_poly1} and~\ref{tab:win_lpo_poly} by denoting window size in ms as $N_t$. 
It can be seen that the performance of the TVQCP methods is quite stable over a range of values for the window size and polynomial order.
However, the performance seems to be slightly sensitive to the choice of prediction order, which needs further investigation.

\subsubsection{The choice of weighting function}
The effect of using different weighting functions within the framework of TVWLP for $L_1$ norm and $L_2$ norm on formant tracking performance is given in Table~\ref{tab:wt_fn1} and ~\ref{tab:wt_fn}.
The different weighting functions studied include the signal energy--based STE function, the residual--based weighting function, and the QCP weighting function discussed earlier in Section~\ref{sec:wt_fn}. It can be seen that the QCP weighting function performs best among the three compared weighting functions. Note that the TVWLP method with the QCP weighting in Table~\ref{tab:wt_fn1} and Table~\ref{tab:wt_fn} corresponds to TVQCP-L1 analysis and TVQCP-L2 analysis, respectively.

\subsubsection{Robustness to GCI detection errors and the DQ parameter}
\label{Robustness_to_GCI_DQ}
The robustness of the proposed TVQCP method to errors in GCI detection was studied by artificially inducing errors in the estimated GCI locations.
Two types of errors were studied.
In the first, a uniformly distributed random error ($Rerr$) was added to the estimated GCIs.
In the second, there was a fixed error ($Ferr$) that gives a consistent bias to the estimated GCIs.
The formant tracking results for random and fixed GCI errors is given in Table~\ref{tab:gci1} and~\ref{tab:gci} for TVQCP-L1 and TVQCP-L2, respectively.
It can be seen that the performance of the proposed TVQCP methods is robust to GCI errors in the range of 1--2 ms.

Simulating a fixed GCI error is equivalent to altering the position quotient (PQ) of the QCP weighting function (described in Section~\ref{sec:qcpwt}). The performance of TVQCP in relation to varying the duration quotient (DQ) of the QCP weighting function between 0.5 and 0.9 is given in Tables~\ref{tab:gci1} and~\ref{tab:gci} using $L_1$ and $L_2$ norm minimization, respectively. It can be seen that TVQCP performed robustly over a range of DQ values, and the best performance was obtained with DQ=0.8, i.e., using a weighting function, which suppresses the residual energy in 20\% of the samples during the fundamental period. Therefore, this value of DQ was used in all the analyses of the study.

\begin{table}[t]
\centering
\caption{\label{tab:compare} Formant tracking performance on natural speech data in terms of FDR and FEE for different methods.}
\vspace{-0.1cm}
{\small
\begin{tabular}{|c||c|c|c||c|c|c|}
\hline
 & \multicolumn{3}{|c||}{FDR (\%)} & \multicolumn{3}{|c|}{FEE (Hz)} \\\cline{2-7}
Method & $F_1$ & $F_2$ & $F_3$ & $\delta F_1$ & $\delta F_2$ & $\delta F_3$\\\hline\hline
PRAAT & 86.0  & 70.0  & 63.1  & 87.9 & 268.3 & 340.1 \\\hline 
MUST & 81.1  & 86.3  & 76.9  & 90.5 & 152.3 & 229.8 \\\hline 
WSURF & 86.6  & 82.7  & 80.8  & 87.3 & 222.5 & 228.2 \\\hline 
KARMA &  91.5 & 89.4 & 74.7 & {\bf 61.9} & 145.8 & 250.3 \\\hline 

DeepF & {\bf91.7} & 92.3 & 89.7 & 85.1 & 119.6 & 142.8 \\\hline 
TVLP-L2 & 88.8 & 94.9 & 89.3 & 70.1 & 104.9 & 149.8 \\\hline
TVQCP-L2 & 90.6 & 96.1 & 92.0 & 68.4 & 94.5 & 126.2 \\\hline
TVLP-L1  &90.3 & 96.2 & 91.7 & 68.9 & 94.4 & 129.6  \\\hline
TVQCP-L1 & 90.9 & {\bf96.5} & {\bf92.2} & 67.6 & {\bf91.7} & {\bf123.9} \\\hline
\end{tabular}
}
\vspace*{-0.2cm}
\end{table}

\begin{table}[t]
\centering
\small\addtolength{\tabcolsep}{-1pt}
\caption{\label{tab:tvqcpvskarma} The formant tracking performance of KARMA, DeepF, TVLP-L1 and TVQCP-L1 in terms of FDR and FEE for different phonetic categories of natural speech data. 
\vspace{-0.2cm}
}
\begin{tabular}{|c||c|c|c||c|c|c|}
\hline
 & \multicolumn{3}{c||}{FDR (\%)} & \multicolumn{3}{c|}{FEE (Hz)} \\\cline{2-7}
Phonetic category~ & ~$F_1$~ & ~$F_2$~ & ~$F_3$~ & ~$\delta F_1$~ & ~$\delta F_2$~ & ~$\delta F_3$\\\hline
\multicolumn{7}{c}{}\\\hline
\multicolumn{7}{|c|}{{\bf KARMA}} \\\cline{2-7} \hline 
Vowels (V)     & { 92.6} & 89.0 & 74.5 & {\bf 57.1} & 149.5 & 251.1  \\\hline
Diphthongs (D) & { 92.5} & 92.3 & 76.5 & {\bf 62.8} & 128.7 & 239.8  \\\hline
Semivowels (S) & { 86.9} & 86.9 & 73.6 & {\bf 76.1} & 154.8 & 258.3   \\\hline
V+D+S      & { 91.5} & 89.4 & 74.7 & {\bf 61.9} & 145.8 & 250.3  \\\hline
All voiced sounds & { 87.9} & 88.4 & 75.0 & {\bf 70.9} & 151.7 & 248.0 \\\hline 
\multicolumn{7}{c}{}\\\hline

 \multicolumn{7}{|c|}{{\bf DeepF}} \\\cline{2-7} \hline
 Vowels (V)     & {\bf92.7} & 93.7 & 91.0 & 81.5 & 112.9 & 135.4  \\\hline
Diphthongs (D) & {\bf93.2} & 93.8 & 90.6 & 84.8 & 112.2 & 132.9  \\\hline
Semivowels (S) & {\bf87.0} & 86.1 & 84.4 & 96.1 & 148.4 & 176.2  \\\hline
V+D+S      & {\bf91.7} & 92.3 & 89.7 & 85.1 & 119.6 & 142.8  \\\hline
All voiced sounds &{\bf88.8} & 90.6 & {\bf88.7} & 86.8 & {\bf129.7} & {\bf147.4}  \\\hline 
\multicolumn{7}{c}{}\\\hline
 \multicolumn{7}{|c|}{{\bf TVLP-L1}} \\\cline{2-7} \hline
Vowels (V)  &91.9 & 97.0 & 92.8 & 63.3 & 86.4 & 117.8  \\\hline
Diphthongs (D)  &92.0 & 98.1 & 93.7 & 64.9 & 81.4 & 114.9 \\\hline
Semivowels (S)  &83.6 & 91.6 & 85.9 & 90.1 & {\bf133.4} & {\bf182.0}  \\\hline
V+D+S      &90.3 & 96.2 & 91.7 & 68.9 & 94.4 & 129.6  \\\hline
All voiced sounds &83.6 & 90.9 & 87.2 & 96.0 & 133.8 & 162.9 \\\hline 
\multicolumn{7}{c}{}\\\hline
 \multicolumn{7}{|c|}{{\bf TVQCP-L1}} \\\cline{2-7} \hline
Vowels (V)  & 92.6 & {\bf97.4} & {\bf93.6} & 62.0 & {\bf82.5} & {\bf110.6}  \\\hline
Diphthongs (D) & 92.5 & {\bf98.3} & {\bf94.3} & 63.6 & {\bf77.4} & {\bf107.7}  \\\hline
Semivowels (S) & 84.2 & {\bf91.7} & {\bf85.5} & 89.0 & 135.7 & 182.5  \\\hline
V+D+S      & 90.9 & {\bf96.5} & {\bf92.2} & 67.6 & {\bf91.7} & {\bf123.9} \\\hline
All voiced sounds & 84.1 & {\bf91.2} & 87.5 & 95.1 & 131.4 & 159.2  \\\hline 
\end{tabular}
\end{table}


\subsubsection{A comparison of time-variant linear predictive methods and other formant tracking methods for clean speech}
The performance of the TVLP and TVQCP methods with different norms are compared to some of the popular formant tracking methods in Table~\ref{tab:compare}. 
``PRAAT" denotes the Burg method of LP analysis with a 50-ms Gaussian-like window function that is used in formant tracking in Praat, a widely used speech research tool~\cite{praat2001}. ``MUST" denotes an adaptive filter-bank based method proposed by Mustafa et al.~\cite{Mustafa2006}.
``WSURF" denotes the formant tracker part of Wavesurfer~\cite{wavesurfer2000} that uses a stabilized covariance analysis over a 25-ms Hamming window.
``KARMA" denotes the state-of-the-art KF--based formant tracking method published in~\cite{Mehta2012}. ``DeepF" (DeepFormants) denotes the deep-learning based formant tracking method proposed recently in ~\cite{Dissen2016,Dissen2019,DeepFwebsite}. It is worth emphasizing that DeepF is based on supervised learning and calls for an annotated speech corpus to be trained.

\begin{table}
\centering
\caption{\label{tab:noisecompare} Formant tracking performance for different methods using speech degraded with volvo, babble and white noise at SNR levels of 20 dB, 10 dB, and 5 dB.}
\vspace{-0.2cm}
\resizebox{8cm}{8cm}{
\begin{tabular}{|c||c|c|c||c|c|c|}
\hline
 & \multicolumn{3}{|c||}{FDR (\%)} & \multicolumn{3}{|c|}{FEE (Hz)} \\\cline{2-7}
Method & $F_1$ & $F_2$ & $F_3$ & $\delta F_1$ & $\delta F_2$ & $\delta F_3$\\\hline

\multicolumn{7}{c}{}\\\hline
\multicolumn{7}{|c|}{{\bf Volvo at 20 dB}} \\\hline
KARMA       & 90.1 & 88.5 & 73.3 & 68.1 & 153.8 & 266.4  \\\hline
DeepF       & 90.0 & 92.0 & 88.2 & 96.1 & 117.5 & 149.3   \\\hline
TVQCP-L1    & {\bf91.0} & {\bf96.2} & {\bf92.1} & {\bf68.0} & {\bf93.2} & {\bf125.3}  \\\hline

\multicolumn{7}{|c|}{{\bf Volvo at 10 dB}} \\\hline
KARMA        &86.2 & 86.6 & 71.7 & 80.4 & 167.6 & 278.8  \\\hline
DeepF        &89.4 & 91.5 & 87.4 & 97.3 & 120.4 & 153.7   \\\hline
TVQCP-L1     &{\bf90.7} & {\bf95.8} & {\bf91.5} & {\bf70.9} & {\bf96.9} & {\bf130.2}\\\hline

\multicolumn{7}{|c|}{{\bf Volvo at 5 dB}} \\\hline
KARMA        &80.8 & 84.9 & 70.6 & 96.4 & 182.9 & 299.3  \\\hline
DeepF        &89.7 & 90.7 & 85.9 & 95.4 & 124.7 & 160.3   \\\hline
TVQCP-L1     &{\bf89.8} & {\bf95.1} & {\bf90.8} & {\bf76.3} & {\bf105.2} & {\bf138.2}   \\\hline 

\multicolumn{7}{c}{}\\\hline
\multicolumn{7}{|c|}{{\bf Babble at 20 dB}} \\\hline
KARMA        &{\bf91.7} & 88.0 & 74.2 & {\bf61.4} & 152.6 & 247.8  \\\hline
DeepF        &91.3 & 91.7 & 87.1 & 89.7 & 118.0 & 155.1   \\\hline
TVQCP-L1     &89.6 & {\bf94.7} & {\bf89.9} & 68.6 & {\bf103.2} & {\bf136.9}  \\\hline

\multicolumn{7}{|c|}{{\bf Babble at 10 dB}} \\\hline
KARMA        &90.3 & 83.8 & 71.8 & {\bf65.1} & 176.1 & 246.0  \\\hline
DeepF        &{\bf91.1} & 86.6 & 81.7 & 88.4 & 145.9 & 182.7   \\\hline
TVQCP-L1     &84.3 & {\bf88.1} & {\bf82.5} & 78.1 & {\bf144.5} & {\bf181.0} \\\hline

\multicolumn{7}{|c|}{{\bf Babble at 5 dB}} \\\hline
KARMA        &88.2 & 78.9 & 68.7 & {\bf70.9} & 200.9 & 260.3   \\\hline
DeepF        &{\bf89.8} & 81.4 & 76.1 & 89.9 & 177.3 & {\bf209.1}   \\\hline
TVQCP-L1     &80.9 & {\bf83.2} & {\bf76.6} & 86.4 & {\bf174.0} & 212.7 \\\hline

\multicolumn{7}{c}{}\\\hline

\multicolumn{7}{|c|}{{\bf White at 20 dB}} \\\hline
KARMA        &90.4 & 87.6 & 73.6 & {\bf64.4} & 150.5 & 240.6 \\\hline
DeepF        &90.1 & 90.4 & {\bf84.4} & 95.4 & {\bf125.9} & {\bf167.9}   \\\hline

TVQCP-L1     &{\bf92.0} & {\bf93.0} & 79.6 & 68.3 & 129.1 & 205.7   \\\hline

\multicolumn{7}{|c|}{{\bf White at 10 dB}} \\\hline
KARMA        &86.2 & 80.1 & 68.8 & 75.5 & 191.3 & 256.5 \\\hline
DeepF        &89.8 & 80.8 & {\bf71.6} & 99.2 & 184.3 & {\bf238.7}   \\\hline
TVQCP-L1     &{\bf90.3} & {\bf85.2} & 66.1 & {\bf73.7} & {\bf179.2} & 280.7 \\\hline

\multicolumn{7}{|c|}{{\bf White at 5 dB}} \\\hline
KARMA        &80.1 & 72.5 & 64.0 & {91.6} & 232.5 & 279.2\\\hline
DeepF        &{\bf89.2} & 71.7 & {\bf64.5} & 101.1 & 238.7 & {\bf274.3}  \\\hline
TVQCP-L1     & 88.3 & {\bf77.9} & 59.6 & {\bf82.4} & {\bf220.1} & 314.5 \\\hline
\end{tabular}
}
\end{table}

It can be seen from Table~\ref{tab:compare} that the TVLP and TVQCP methods clearly performed better (a 20--60\% reduction in error across the three formants) compared to the popular formant tracking methods (Praat and Wavesurfer) that use a two-stage detect-and-track approach.
The proposed TVQCP method provided an improvement in the performance (both FDRs and FEEs) of tracking  the second and third formants (a reduction in the estimation error of around 30\% and 50\% respectively) compared to KARMA. The KARMA method performed slightly better than the TVQCP method (with a relative improvement of around 9\%) in tracking the first formant. Compared to DeepF, the proposed TVQCP method provided an improvement in FEEs of around 20\%, 21\% and 12\% for all the three formants, respectively.
In terms of FDR, DeepF performed slightly better (around 1\%) than TVQCP for the first formant. However, for the second and third formants, TVQCP improved the FDR by around 4\% and 3\%, respectively, compared to DeepF.
Differences in performance within the family of time-varying methods were not as evident.
However, it can be seen from the results that the use of TVQCP analysis seems to improve the performance of formant tracking. It can also be observed that TVLP-L1 is slightly better than TVLP-L2, and TVQCP-L1 is slightly better than TVQCP-L2. Between TVLP-L1 and TVQCP-L1, TVQCP-L1 is better than TVLP-L1 in both FDRs and FEEs for all the three formants (a reduction in the estimation error of around 2\%, 3\% and 4\% for F1, F2 and F3, respectively).

\begin{figure*}[h!]
\includegraphics[width=18cm]{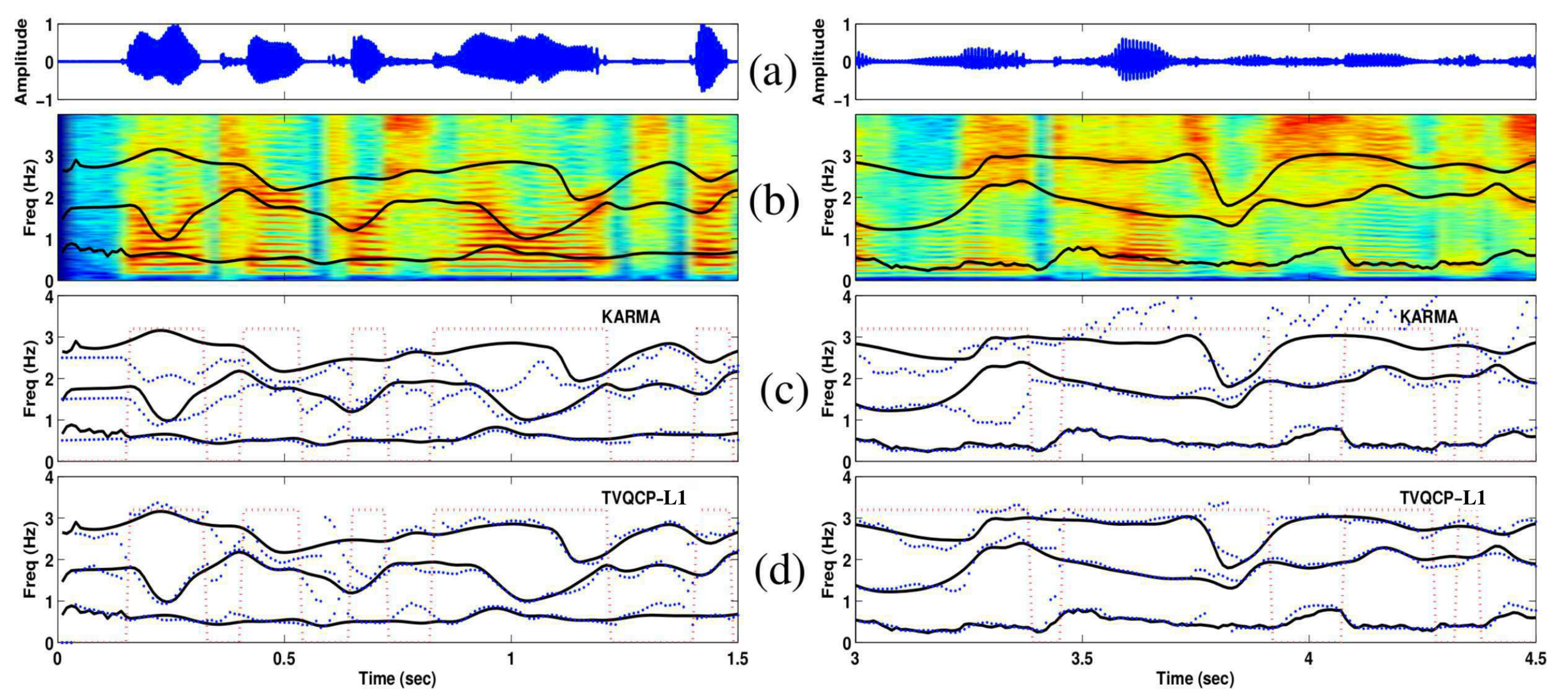} 
\vspace*{-0.3cm}
\caption{\label{fig:tvqcpvskarma} Examples of formant tracks estimated by KARMA and TVQCP-L1 from utterances produced by a female (on the left) and male (on the right) talker: (a) time-domain speech signals, (b) a narrowband spectrogram with reference ground truth formant contours, (c) formant track estimates using KARMA along with the voiced--unvoiced regions shown by a dotted rectangular-wave plot, and (d) formant track estimates using TVQCP-L1.}
\end{figure*}

A detailed comparison in the formant tracking performance of KARMA, DeepF, TVLP-L1 and TVQCP-L1 is given in Table~\ref{tab:tvqcpvskarma} for different phonetic categories.
It can be seen that the estimation error of TVQCP-L1 is 15--40\% and 25--55\% smaller than that of KARMA for $F_2$ and $F_3$ respectively. Likewise, KARMA gave an estimation error that was 1--15\% smaller than that of TVQCP-L1 for $F_1$ across different phonetic categories.
In comparison to DeepF, the estimation error of TVQCP-L1 was 7--25\%, 9--30\% and 13--20\% smaller for $F_1$, $F_2$ and $F_3$, respectively (except in semivowels for $F_3$). The performance of DeepF for $F_3$ in semivowels was better (by around 4\%) than that of TVQCP-L1. It can also be observed that the performance of TVLP-L1 for $F_2$ and $F_3$ in semivowels was slightly better (by around 1\%) than that of TVQCP-L1. On the other hand, the performance of TVQCP-L1 for $F_1$, $F_2$ and $F_3$ was better (by around 2--6\%) than that of TVLP-L1. When all the voiced sounds are considered, the performance of DeepF was better than the other methods reflecting the fact that DeepF benefits from supervised learning of the formant contours in the model training. Note that the reliability of the manually corrected reference ground truth is less  for the other phonetic categories. In view of this, we can argue that the proposed TVQCP method provided the best overall formant tracking performance compared to the popular and state-of-the-art reference methods. A qualitative comparison of the formant tracking performances of the TVQCP-L1 and KARMA methods is demonstrated by Fig.~\ref{fig:tvqcpvskarma} for utterances produced by a male and female speaker. It can be seen from the figure that the TVQCP-L1 method clearly performed better than KARMA in tracking $F_2$ and $F_3$, with a comparable performance for $F_1$.

\subsubsection{A comparison of time-variant linear predictive methods and other formant tracking methods for noisy speech}
In this section, the performance of the TVQCP-L1 method is compared to KARMA and DeepF in formant tracking of noisy speech, as these methods were shown to perform better than the other methods for clean speech. Noisy speech was obtained by  corrupting the original clean signals of the VTR speech database with different types of additive noise. The results obtained for stationary and non-stationary noise of three types (volvo, babble, and white) in three SNR categories (20 dB, 10 dB and 5 dB) are given in Table~\ref{tab:noisecompare}. From the results, it can be observed that the performance of the methods drops as the SNR decreases from 20 dB to 5 dB.
For signals corrupted by volvo noise, it is observed that the performance of TVQCP-L1 is better than that of the other methods in all SNR categories in both FDR and FEE for all the formants. In the case of the babble noise, the methods behave similarly as in clean speech. That is, FDR of $F_1$ is better for DeepF, FEE of $F_1$ is better for KARMA, and the proposed method is better in both FDR and FEE for $F_2$ and $F_3$ formants.
In the case of white noise (without pre-emphasis), the performance of the proposed method for $F_1$ and $F_2$ is better than others in both FDR and FEE (except at 20 dB SNR). Whereas, the performance of the DeepF method for $F_3$ is better than others in both FDR and FEE. 

\section{Conclusions}
\label{sec:summary}
In this paper, we proposed a new formant tracking method, TVQCP\footnote{https://github.com/njaygowda/ftrack}, for speech signals. The TVQCP method combines the advantages of QCP analysis (reducing the effect of the glottal source in formant estimation by using temporal weighting of the prediction error), the increased sparsity of the prediction error due to the $L_1$ norm minimization and TVLP analysis (imposing a time-continuity constraint to take into account the slowness of movements in the real human vocal tract).
The use of a time-continuity constraint on the vocal tract parameters eliminates the need for a two-stage detect-and-track strategy to combine them into one.
Formant tracking experiments on synthetic speech utterances demonstrate the advantages of the proposed TVQCP method over TVLP.
A comparison of performance on natural speech utterances shows that the TVQCP method performs better than some of the state-of-the-art formant trackers, such as Praat, Wavesurfer, KARMA and DeepFormants. One limitation of the proposed TVQCP method is its apparent sensitivity to the choice of prediction order, though a prediction order of 8 works consistently well in tracking formants over a large set of natural speech utterances of male and female talkers.
In addition, there is a need for devising a better coasting strategy (such as the one that is used in KARMA) for tracking formants in non-speech and non-voiced sections or in less-reliable voiced regions.

\section*{Acknowledgements}
The authors would like to thank the authors of Deep Formants for their helpful discussions. 

\begin{biography}[{\includegraphics[width=1.1in,height=1.25in,clip,keepaspectratio]{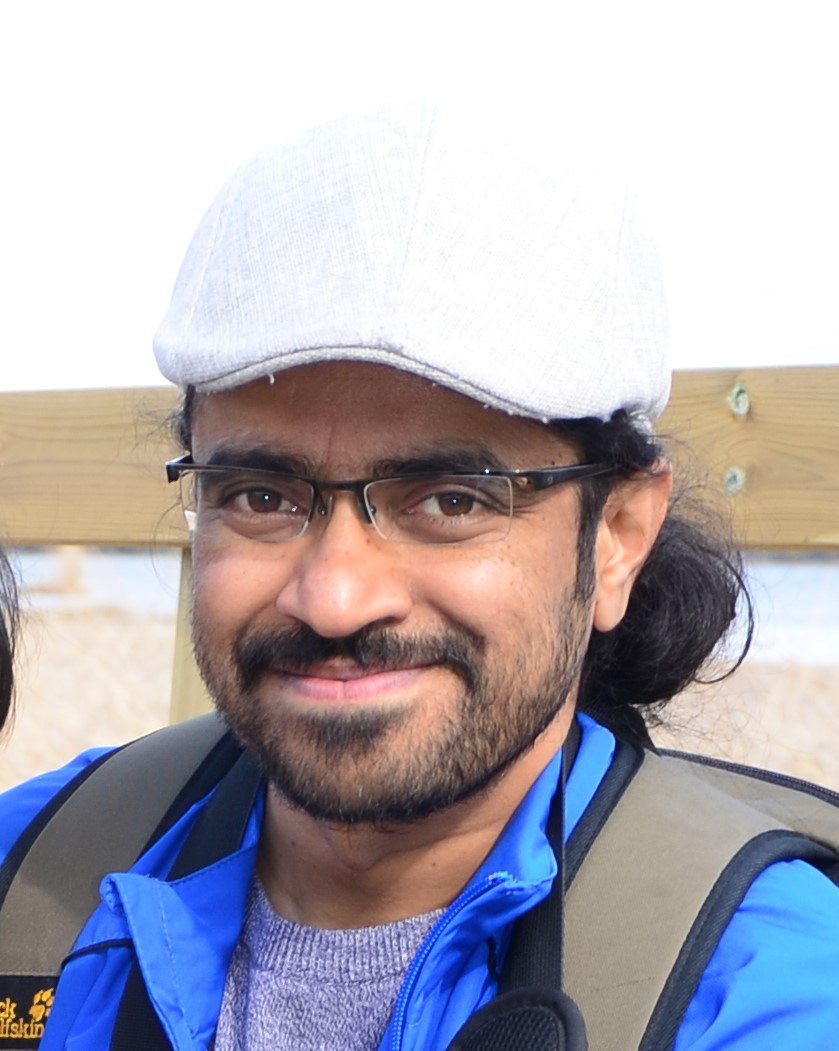}}]{Dhananjaya N. Gowda } is currently working as a Principal Engineer at Speech Processing Lab, AI Center, Samsung Research at Seoul R\&D Campus, South Korea. He worked as a postdoctoral researcher at Aalto University, Espoo, Finland from 2012 to 2017. He holds a Doctorate (2011) and a Master's degree (2004) in the area of speech signal processing, both from the Dept. of Computer Science and Engineering, Indian Institute of Technology (IIT) Madras, Chennai, India. His research interests include speech processing, signal processing, speech recognition, machine learning and spoken language understanding.
\end{biography}

\begin{biography}[{\includegraphics[width=1.1in,height=1.25in,clip,keepaspectratio]{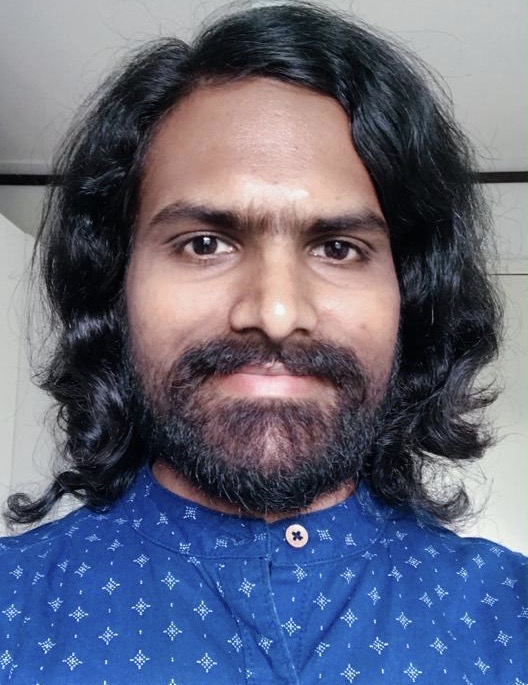}}]{Sudarsana Reddy Kadiri} received Bachelor of Technology degree from Jawaharlal Nehru Technological University (JNTU), Hyderabad, India, in 2011, with a specialization in Electronics and Communication Engineering (ECE), the M.S. (Research) during 2011-2014, and later converted to Ph.D., and received Ph.D. degree from the Department of ECE, International Institute of Information Technology, Hyderabad (IIIT-H), India in 2018. He was a Teaching Assistant for several courses at IIIT-H during 2012-2018. He is currently a Postdoctoral Researcher with the Department of Signal Processing and Acoustics, Aalto University, Espoo, Finland. His research interests include signal processing, speech analysis, speech synthesis, paralinguistics, affective computing, voice pathologies, machine learning and auditory neuroscience.
\end{biography}

\begin{biography}[{\includegraphics[width=1.1in,clip,keepaspectratio]{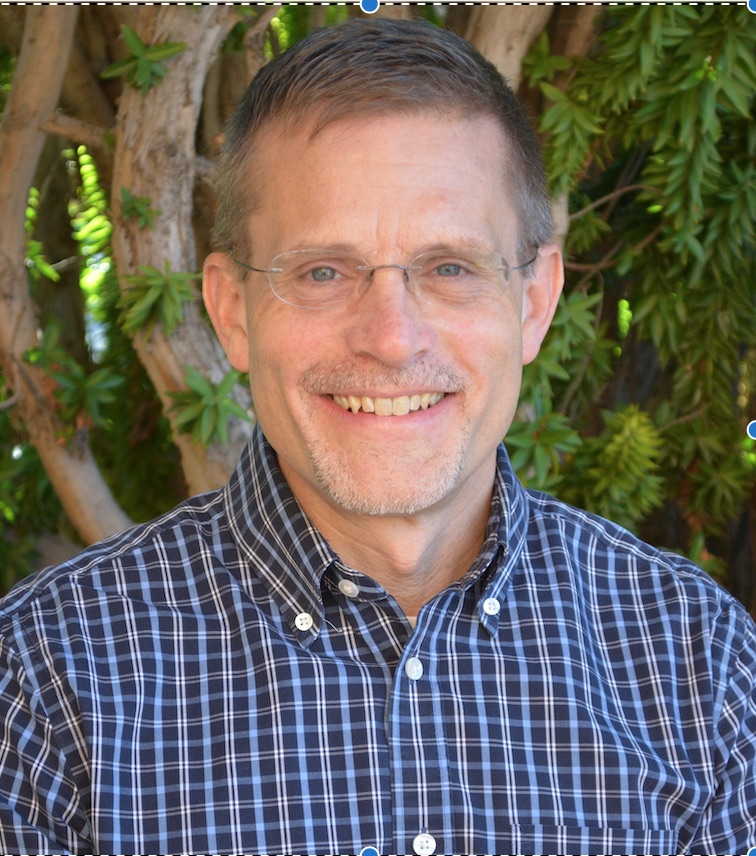}}]{Brad Story} is a Professor in the Department of Speech, Language, and Hearing Sciences at the University of Arizona. His research is concerned with development of computational, physically-based models that simulate the observed structure, movement, and acoustic characteristics of specific components of the speech production system. He has taught courses at both the undergraduate and graduate levels in Speech Science, Speech Perception, Acoustics, Hearing Science, and Anatomy and Physiology. Dr. Story is a fellow of the Acoustical Society of America, recipient of the Rossing Prize in Acoustics Education and Willard R. Zemlin Lecture in Speech Science, and has served as an Associate Editor of the Journal of the Acoustical Society of America. He has authored over 100 publications in the area of voice and speech science.
\end{biography}

\begin{biography}[{\includegraphics[height=1.25in,trim=0cm 0 0cm 0,clip]{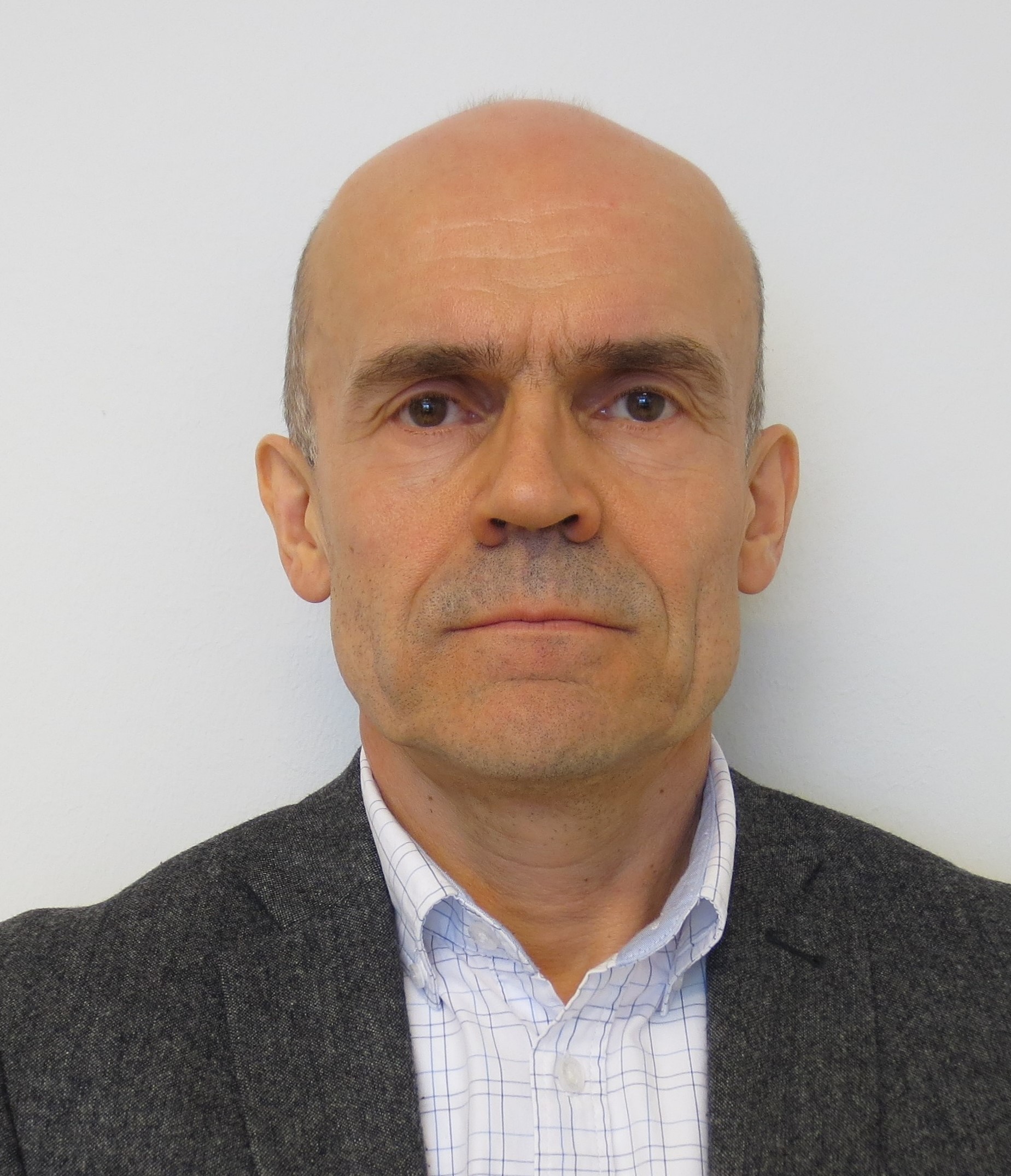}}]{Paavo Alku} received his M.Sc., Lic.Tech., and Dr.Sc.(Tech) degrees from Helsinki University of Technology, Espoo, Finland, in 1986, 1988, and 1992, respectively. He was an assistant professor at the Asian Institute of Technology, Bangkok, Thailand, in 1993, and an assistant professor and professor at the University of Turku, Finland, from 1994 to 1999. He is currently a professor of speech communication technology at Aalto University, Espoo, Finland. His research interests include analysis and parameterization of speech production, statistical parametric speech synthesis, spectral modelling of speech, speech-based biomarking of human health, and cerebral processing of speech. He has published more than 200 peer-reviewed journal articles and more than 200 peer-reviewed conference papers. He is an Associate Editor of J. Acoust. Soc. Am. He served as an Academy Professor assigned by the Academy of Finland in 2015-2019. He is a Fellow of the IEEE and a Fellow of ISCA. 
\end{biography}

\end{document}